# Dificuldades na aprendizagem de Física sob a ótica dos resultados do Enem

*Physics learning difficulties from the perspective of ENEM results*

Marta F. Barroso[a], Gustavo Rubini[b], Tatiana da Silva[c]

[a] *UFRJ/IF, marta@if.ufrj.br*
[b] *UFRJ/PEMAT, gustavorubini@if.ufrj.br*
[c] *UFSC/FSC, tatiana.silva@ufsc.br*

**Resumo**
Os resultados do Exame Nacional do Ensino Médio constituem-se em ferramenta importante para diagnóstico das deficiências do ensino no final de um ciclo formativo, uma fonte relevante de dados para a avaliação do que foi aprendido pelos estudantes ao final do ensino médio. Neste trabalho, apresentam-se treze questões selecionadas das provas de Ciências da Natureza do Enem no período entre 2009 e 2014, classificadas disciplinarmente como "questões de física", nas quais algumas das alternativas de respostas revelam concepções não científicas bem delimitadas e descritas na literatura. A partir do desempenho nessas questões dos estudantes concluintes do ensino médio no ano do exame, fazem-se observações a respeito do processo de aprendizagem em física na educação básica. Esse estudo revela que há algumas dificuldades permanentes na compreensão de conceitos básicos de mecânica, fenômenos térmicos e ótica geométrica. Muitas dessas questões evidenciam a presença e eventual predominância de concepções não científicas. Mesmo com as dificuldades para modificar estas concepções, os resultados revelam que, apesar de todo o esforço desenvolvido de pesquisa desde os anos 1980, houve pouco impacto dos resultados no processo de ensino e aprendizagem, e pequena incorporação deste conhecimento acumulado nos materiais didáticos e nos cursos de formação de professores.

**Palavras-chave:** ensino de física, ENEM, concepções não científicas.

**Abstract**
The results of the National High School Examination (ENEM) are an important tool for a diagnosis of educational deficiencies at the end of a training cycle. This exam is a relevant source of data for the evaluation of what is learned by high school students. In this paper, thirteen selected questions are presented, from the Natural Sciences exam in the period between 2009 and 2014, questions that were classified as "physics questions", in which the alternatives reveal well-delimited non-scientific conceptions described in the literature. From the students' performance on these items, observations about the learning process in physics in basic education are made. This study reveals that there are some permanent difficulties in understanding basic concepts of mechanics, thermal phenomena and geometric optics. Many of these questions point to the presence and even predominant role of non-scientific conceptions described in the research literature in physics teaching. Even knowing how difficult it is to modify these conceptions, the results reveal that, despite all the research effort in physics teaching since the 1980s, there was little impact of the results in the learning process, and only a few changes incorporating this knowlege in the didactic materials and in the teacher training courses.

**Introdução**

Avaliações em larga escala, os grandes exames ao final de um ciclo ou de avaliação de uma etapa de um sistema educacional, não costumam ser apreciadas pelos professores e por pesquisadores em ensino e educação. Esses exames em geral são objeto de críticas tanto em relação a seus objetivos quanto a seus métodos e ao instrumento de avaliação, as questões da prova.

No entanto, observa-se que essas avaliações se tornaram muito frequentes na área de Ciências a partir dos anos 90 [1], com o desenvolvimento do exame internacional de avaliação de estudantes na faixa de 15 anos – PISA, que se propunha a avaliar o denominado letramento científico para os países da comunidade européia e convidados [2]. Os avanços técnicos nas possibilidades de realização, correção e apresentação de notas comparáveis para estas provas, devido aos desenvolvimentos na psicometria com a adoção da Teoria da Resposta ao Item [3], associados às facilidades introduzidas pelos desenvolvimentos tecnológicos em computação, tornaram essas provas muito comuns.

Há poucos trabalhos de pesquisa relativos ao impacto das avaliações externas, em particular na atuação de professores [4]. A utilização dos resultados dessas avaliações que envolvem políticas de responsabilização [5] partem da hipótese que os professores são capazes de utilizar os dados gerados pelas avaliações externas [4].

O Exame Nacional do Ensino Médio, o Enem, pode ser considerado uma avaliação em larga escala com aspectos peculiares. Criado em 1998, propunha-se a fornecer um mecanismo de auto-avaliação para o estudante ao final do ensino médio. Em 2008, seus objetivos, métodos e práticas foram profundamente alterados [6], em função da necessidade de criação de um sistema de classificação para ingresso no ensino superior público e para concessão de bolsas e financiamento para o ensino superior privado. A prova generalista foi substituída por provas objetivas em quatro áreas (Ciências da Natureza, Ciências Humanas, Matemática e Linguagens).

Esse exame não pode ser facilmente classificável como uma avaliação em larga escala, pois não é universal (já que o exame não é obrigatório para todos os alunos que concluem o ensino médio) e também não é amostral. No entanto, por se constituir em um caminho para o ingresso no ensino superior, seja por meio do Sistema de Seleção Unificada (SiSU) que classifica os estudantes para ingresso em instituições federais de ensino superior, seja por meio dos editais para bolsas e financiamentos para o ensino superior privado, torna-se um dos exames mais valiosos para o estudo da aprendizagem dos estudantes ao final da educação básica – é um exame que vale muito para quem o presta, e neste sentido os resultados revelam em geral o melhor que os alunos podem fazer. Essa importância pode ser verificada pelo fato que o número de estudantes que completam o ensino médio a cada ano de 2009 a 2012 é praticamente constante, da ordem de 1,8 milhões [7], e o número de estudantes concluintes do ensino médio que faz o Enem varia entre 0,8 e 1,4 milhões [8] no período entre 2009 e 2014.

Os resultados do Enem, nestas condições, possibilitam o estudo do currículo aprendido na educação básica e a comparação com as prescrições curriculares (o currículo oficial) e o que é ensinado. E esta é uma ferramenta crucial para um diagnóstico claro das deficiências do ensino. Pelas características do exame, espera-se que o estudante tenha atingido algum nível de compreensão esperado para um jovem adulto; entretanto, de acordo com Anderson [9],

> *"Os pesquisadores em ensino de ciências também estão, em geral, de acordo com um resultado central relativo à prática corrente na escola: <u>nossas instituições de educação formal não ajudam a maioria dos estudantes a aprender ciências com compreensão.</u> Este é um resultado robusto (...)"*.[1] *(pág.5)*

---

[1] Versão livre do original, *"researchers on science education also generally agree on one central finding about current school practice: Our institutions of formal education do not help most students to learn science with understanding. This is a robust finding (...)"* (p. 5)

A partir do estudo dos itens e do desempenho dos alunos nas questões de Física do Enem [6;8;10;11;12] podem ser obtidas evidências relativas à compreensão dos conceitos de física pelos estudantes ao final da educação básica. Esse estudo revela que há algumas dificuldades permanentes na compreensão de conceitos, mais do que aspectos sobre o desenvolvimento de competências e habilidades, ou mesmo de dificuldades algébricas ou matemáticas. Muitas dessas questões evidenciam a presença forte, no final do ensino médio, de concepções não científicas descritas na literatura de pesquisa em ensino de física já há várias décadas.

Neste trabalho, apresentam-se questões selecionadas das provas do Enem no período entre 2009 e 2014 nas quais as alternativas de respostas (denominadas distratores, quando não correspondem ao gabarito) revelam concepções não científicas bem delimitadas e descritas na literatura, e a partir do desempenho dos estudantes nessas questões fazem-se observações a respeito do processo de aprendizagem em física na educação básica.

**Distratores relacionados a concepções não científicas**

No período entre os anos 1970 e 1990, pesquisadores em ensino de física deram-se conta da necessidade de compreender o que os alunos traziam como bagagem cultural e conceitual para que pudessem desenvolver processos de aprendizagem em física. Essa tomada de consciência tornou-se mais importante pela percepção de existência de um conjunto de "concepções alternativas", não fazendo parte do corpo de conceitos considerados científicos, que a maior parte das crianças e adolescentes possuem, em todo o mundo, e que os municiam com explicações próprias para compreensão do mundo natural.

Os estudantes chegam à sala de aula com explicações próprias a respeito dos fenômenos que ocorrem ao seu redor. A existência de concepções desenvolvidas espontaneamente a respeito de fenômenos físicos é objeto de extensa literatura de pesquisa [13; 14]. Segundo Zylberstajn [13],

> *"Pesquisas têm demonstrado que estas concepções, na forma de expectativas, crenças, princípios intuitivos, e significados atribuídos a palavras cobrem uma vasta gama dos conceitos que fazem parte dos currículos de disciplinas científicas. É igualmente verdadeiro que, para muitos, algumas destas noções são fortemente incorporadas à sua estrutura cognitiva, tornando-se resistentes à instrução."*

Esses trabalhos evidenciam as dificuldades de estudantes e professores em diferentes níveis de ensino com os conceitos básicos de física. Apesar desses resultados conhecidos, há poucas mudanças nos materiais didáticos e nos cursos de formação de professores que incorporem esse conhecimento; aparentemente, não chegam à sala de aula [15].

Esta linha de atuação na pesquisa em ensino fundamentou-se e evoluiu, mas a compreensão, conhecimento e pesquisa relativa às ideias prévias, em sua maior parte não escolarizadas, dos estudantes continuou a fazer parte de um dos principais paradigmas de pesquisa da área [9]. Há um conjunto dessas concepções organizadas e catalogadas; o trabalho de Pfund e Duit [16] é um dos exemplos da compilação dessas concepções.

Nas provas do Enem, é possível observar algumas questões que apresentam, dentre as alternativas de resposta, algumas que seriam escolhidas por estudantes que mantêm, após a instrução formal do ensino médio, as mesmas concepções não científicas relatadas na literatura. A análise do desempenho dos estudantes concluintes do ensino médio no ano de realização do Enem nas questões em que há alternativas em que essas concepções estão presentes pode revelar a sua persistência e fornecer um diagnóstico da aprendizagem em física ao final deste ciclo formativo.

Marcom e Kleinke [17] fazem uma análise de "questões de física" classificadas em um dos denominados objetos de conhecimento da Matriz de Referência do exame [18], "o movimento, o equilíbrio e a descoberta de leis físicas" da prova de Ciências da Natureza e suas Tecnologias para alunos de escolas estaduais concluintes do ensino médio dos anos de 2009 a 2012.

Neste trabalho, apresenta-se uma análise das alternativas incorretas (os distratores) das questões de física da prova de Ciências da Natureza para os estudantes concluintes do ensino médio entre os anos 2009 a 2014 para os objetos de conhecimento elencados na Matriz de Referência. Não são incluídas neste trabalho as questões classificadas como correspondendo ao objeto de conhecimento "Fenômenos elétricos e magnéticos", que, devido a especificidades, serão tratadas posteriormente.

**Como os resultados foram obtidos**

As respostas de cada estudante às questões do Enem são disponibilizadas publicamente[2] pelo Instituto Nacional de Pesquisas Educacionais Anísio Teixeira (Inep), juntamente com as respostas dadas aos questionários com dados pessoais e da escola em que cursou a educação básica. Essas informações são apresentadas na forma de arquivos de texto grandes, que devem ser utilizados com auxílio de programas de bancos de dados. A descrição de como esses dados são preparados para obtenção dos resultados aqui mostrados está disponível em Barroso, Massunaga e Rubini [8; 19].

O banco de dados é preparado de forma que fique disponível, para cada um dos estudantes (sem identificação) que realizou a prova do Enem do ano considerado, todas as respostas relativas a informações pessoais e da escola, bem como as alternativas escolhidas em cada uma das questões da prova de Ciências da Natureza. Na Tabela 1 a seguir, apresentam-se os números envolvidos nestes bancos de dados: os inscritos na prova, os participantes (compareceram ao primeiro dia) e os concluintes (do ensino médio que fizeram todas as provas com a redação válida).

**Tabela 1.** Número de estudantes participando do Enem de 2009 a 2014: inscritos, participantes e concluintes do ensino médio no ano do exame

|  | Enem 2009 | Enem 2010 | Enem 2011 | Enem 2012 | Enem 2013 | Enem 2014 |
|---|---|---|---|---|---|---|
| Inscritos | 4.148.721 | 4.626.094 | 5.380.856 | 5.791.065 | 7.173.563 | 8.722.248 |
| Participantes | 2.330.534 | 3.101.455 | 3.670.241 | 3.942.639 | 4.908.306 | 5.633.954 |
| Concluintes | 865 | 1.059.227 | 1.174.429 | 1.205.063 | 1.326.681 | 1.374.821 |

Neste trabalho, são utilizadas as respostas de todos os estudantes que se declaram concluintes do ensino médio no ano do exame e que participaram de todas as provas (nos dois dias de exame, as provas de Ciências Humanas e Ciências da Natureza no primeiro dia, e as provas de Linguagens e Matemática no segundo dia) e tiveram a redação considerada válida.

---

[2] Na página http://inep.gov.br/microdados

As questões da prova de Ciências da Natureza foram classificadas por disciplina (Física, Biologia e Química) de acordo com o critério proposto por Gonçalves Jr [10]: a questão é de Física quando o conhecimento necessário para encontrar a resposta correta (gabarito) pertence aos objetos de conhecimento de Física da Matriz de Referência do Enem [18]. Essa classificação foi elaborada de forma independente por dois ou mais pesquisadores, e eventuais divergências (que não superaram 3 questões dentre 45 por prova em todos os anos) foram resolvidas por discussão.

As questões consideradas neste trabalho como contendo alternativas incorretas que revelam a permanência de concepções não científicas foram escolhidas a partir de análise independente pelos autores, seguida de discussão até estabelecimento de proposta consensuada.

Os resultados são apresentados indicando o ano da prova e o número da questão. Por exemplo, 2011_Q61 refere-se à questão 61 da prova de Ciências da Natureza de cor azul do exame do ano de 2011.

Os dados relativos à estatística descritiva da prova e das questões (médias globais, total de acertos, percentual de escolha de cada disciplina, entre outros) foram obtidos com a utilização de programas estatísticos (SPSS e R). Os dados relativos à utilização dos parâmetros da Teoria da Resposta ao Item serão apresentados em seguida.

**Elementos da Teoria da Resposta ao Item utilizados**

A Teoria da Resposta ao Item (TRI) é um modelo proveniente da Psicometria, ou teoria dos testes, que foi formulada nos anos 50 como alternativa à denominada Teoria Clássica dos Testes – a que estabelece como nota de uma prova a soma de acertos, entre outras características [3].

A TRI parte de dois postulados básicos [3]:

> *"(a) O desempenho de um respondente em um item de um teste pode ser previsto (ou explicado) por um conjunto de fatores denominados traços, ou traços latentes, ou habilidades; e (b) a relação entre o desempenho no item de um respondente e o conjunto de traços que possibilita o desempenho no teste pode ser descrito por uma função monotonamente crescente denominada função ou curva característica do item (CCI)."*[3]
> (p.7)

A partir desses postulados, são formulados modelos matemáticos para descrever a aptidão dos respondentes e as curvas características dos itens que compõem o teste. Uma das hipóteses presentes nos modelos usuais da TRI é que apenas uma habilidade é medida em cada um dos itens que compõem o teste. No caso do Enem, cada item corresponde a uma das habilidades da Matriz de Referência.

---

[3] Versão livre do original, *"(a) The performance of an examinee on a test item can be predicted (or explained) by a set of factors called traits, latent traits, or abilities; and (b) the relationship between examinees` item performance and the set of traits underlying item performance can be described by a monotonically increasing function called an item characteristic function or item characteristic curve (ICC)."*

Os modelos matemáticos mais utilizados da TRI são os modelos em que a curva característica do item é ajustada com a utilização de um parâmetro, dois parâmetros ou três parâmetros. No caso do Enem, o modelo utilizado é o modelo logístico de três parâmetros, que descreve a curva característica de cada item de acordo com a função

$$P_i(\theta) = c_i + (1 - c_i)\frac{1}{1 + \exp[-a_i(\theta - b_i)]} \, , \quad i = 1, 2, \cdots, n$$

onde $P_i(\theta)$ é a probabilidade de acerto no item $i$ de um respondente qualquer com aptidão $\theta$, $a_i$ é denominado discriminação do item, e é proporcional à inclinação da curva característica no seu ponto de inflexão, $b_i$ é o parâmetro de dificuldade do item, correspondendo ao ponto de inflexão da curva e $c_i$ é denominado parâmetro de acerto casual (ou pseudo-azar). A curva característica do item $P_i(\theta)$ é uma curva com formato de S com valores entre 0 e 1 sobre toda a escala de aptidão dos respondentes [3].

Na Figura 1, estão traçados exemplos de curvas características de itens com alguns parâmetros diferentes, para identificação das características dessas curvas. Para o estudo do desempenho dos estudantes no Enem, as curvas para os itens serão traçadas, e observados os parâmetros, particularmente de dificuldade e discriminação.

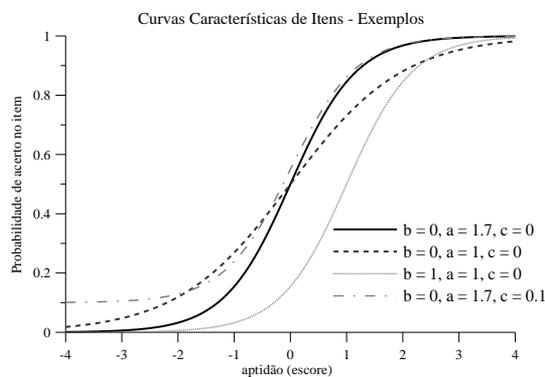

**Figura 1.** Exemplos de curvas características dos itens no modelo logístico de 3 parâmetros, com quatro conjuntos diferentes de parâmetros de dificuldade (b), discriminação (a) e acerto casual (c).

Para traçar a curva característica dos itens a partir do modelo, não se pode contar com as informações do Inep, pois os parâmetros dos itens não são disponibilizados publicamente. No entanto, podem ser utilizados programas, no ambiente estatístico R, que fornecem as informações sobre os escores dos concluintes e os parâmetros dos itens da prova. Os parâmetros aqui apresentados foram obtidos a partir da utilização, no ambiente estatístico R, do pacote mirt [20]. A utilização deste método para ajuste do modelo aos dados forneceu coeficientes de correlação entre os resultados para os escores obtidos e os escores fornecidos pelo Inep superiores a 0,995 em todos os anos estudados.

Em todas as questões a serem discutidas, são apresentados dois conjuntos de curvas. O primeiro é a curva característica do item (CCI), e nela são indicados tanto a curva que representa o ajuste feito a partir da utilização do modelo logístico de 3 parâmetros (a linha contínua) quanto a curva empírica, obtida a partir da divisão dos escores em 20 faixas de igual número de participantes, com a obtenção do percentual de marcação dentro de cada faixa (os pontos).

O segundo conjunto de curvas representa os percentuais de escolha de cada uma das alternativas em função da nota do aluno (dividindo os escores em faixas de 20 pontos).

A curva que descreve o percentual de escolha dos distratores em função da faixa de notas deve ter um padrão complementar ao da curva do gabarito; a probabilidade de erro na questão $Q(\theta)$ corresponde a $Q(\theta) = 1 - P(\theta)$. As curvas da probabilidade de erro devem ter portanto inclinação decrescente.

**Apresentação dos resultados**

As questões das provas de Ciências da Natureza do Enem de 2009 a 2014 foram classificadas disciplinarmente. Dentre as questões de Física desta prova, algumas apresentam, dentre os distratores (as alternativas incorretas), respostas que indicam a persistência de uma concepção não científica entre os alunos.

As questões apresentadas aqui serão nomeadas em função do ano e de sua numeração na "prova azul". Por exemplo, 2009_Q27, refere-se à questão 27 da prova azul de 2009.

Na Tabela 2, apresenta-se o quadro das questões escolhidas para discussão, agrupadas por tema. Na primeira coluna está a informação sobre a questão; na segunda coluna, um "lembrete" sobre o que é abordado na questão; na terceira coluna, o objeto de conhecimento da Matriz de Referência [18] abordado na questão. Os objetos de conhecimento estão indicados por meio de um número de acordo com a sua apresentação na Matriz, como listado:

> *1. Conhecimentos básicos e fundamentais*
> *2. O movimento, o equilíbrio e a descoberta de leis físicas*
> *3. Energia, trabalho e potência*
> *4. A mecânica e o funcionamento do universo*
> *5. Fenômenos elétricos e magnéticos*
> *6. Oscilações, ondas, ótica e radiação*
> *7. O calor e os fenômenos térmicos*

Na coluna seguinte, faz-se referência ao conceito físico abordado na discussão. As colunas posteriores informam o percentual de acertos na questão, o parâmetro de discriminação do modelo ($a$), o parâmetro de dificuldade ($b$) e o parâmetro de acerto casual ou pseudo-azar ($c$).

Desta Tabela 2, observa-se que o percentual de acerto da maioria das questões é baixo, em geral inferior a 20%. O parâmetro de acerto casual também é baixo, e as questões são difíceis (o parâmetro de dificuldade é superior a 500 pontos, que é o valor médio dos escores, entre os concluintes da prova de 2009, e não corresponde a acertar 50% das questões).

**Tabela 2:** A lista das questões analisadas

| Ano_Questão | Lembrete | Obj. Conh. | Conceito envolvido | % acerto | Discriminação (a) | Dificuldade (b) | Acerto Casual (c) |
|---|---|---|---|---|---|---|---|
| 2009_Q27 | Astronauta | 4 | Gravidade | 13,9 | -0,004 | -341,7 | 0,11 |
| 2012_Q60 | Metrô | 2 | Cinemática | 24,4 | 0,056 | 646,8 | 0,23 |
| 2013_Q87 | Paraquedas | 2 | Força e movimento | 15,2 | -0,012 | 201,4 | 0,10 |
| 2014_Q67 | Plano inclinado | 2 | | 14,2 | 0,057 | 662,8 | 0,13 |
| 2014_Q82 | Mônica | 2 | | 12,2 | 0,035 | 696,9 | 0,11 |
| 2010_Q81 | Escultura | 2 | Empuxo | 31,6 | 0,017 | 634,1 | 0,22 |
| 2013_Q76 | Atrito | 2 | Atrito | 19,9 | -0,015 | 164,1 | 0,19 |
| 2012_Q47 | Pneu | 2 | Força e pressão | 43,9 | 0,026 | 536,7 | 0,23 |
| 2013_Q61 | Elevador Hidráulico | 2 | | 27,6 | 0,066 | 650,3 | 0,27 |
| 2010_Q50 | Temperatura e calor | 7 | Temperatura e calor | 18,3 | 0,031 | 713,0 | 0,17 |
| 2013_Q48 | Garrafa PET | 7 | Emissão e absorção de radiação | 18,3 | 0,015 | 689,8 | 0,13 |
| 2011_Q80 | Matriz energética | 7 | Hidrelétrica fonte limpa | 21,4 | 0,020 | 722,2 | 0,20 |
| 2012_Q64 | Pesca | 6 | Raio de luz | 27,6 | 0,034 | 575,1 | 0,16 |

**As concepções sobre gravidade e a questão 27 da prova de 2009**

A pesquisa em ensino de física e de astronomia apresenta muitos estudos realizados desde a década de 70 que apontam que estudantes das mais diferentes faixas etárias e professores possuem concepções não científicas sobre gravidade, massa, peso e força gravitacional [13;21;22] e que podem se assemelhar às ideias de Aristóteles sobre a queda dos corpos [23]. A compreensão destes conceitos está muitas vezes relacionada com o entendimento da forma da Terra e com uma visão espacial da mesma [24;25]. É comum as pessoas considerarem que objetos de massas distintas apresentam tempos de queda diferentes quando soltos de uma mesma altura em virtude das massas serem diferentes, o mais pesado chegando antes ao solo. No lançamento vertical de um objeto, alguns consideram que a força da gravidade só atua na descida. Outros acreditam que ela aumenta com a altura e há ainda aqueles que pensam ser necessário um meio para a transmissão da interação.

A questão 27 da prova azul de 2009, apresentada como consta na prova na Figura 2, aborda alguns destes conceitos. O estudante deve se posicionar em relação a uma afirmação de um astronauta que ao, se aproximar do telescópio Hubble, abre no espaço a porta do ônibus espacial e diz: "*Esse telescópio tem massa grande, mas o peso é pequeno*". Segundo Bassalo [26], a questão apresenta problemas em sua redação, o que, no entanto, não prejudica o estudo aqui apresentado.

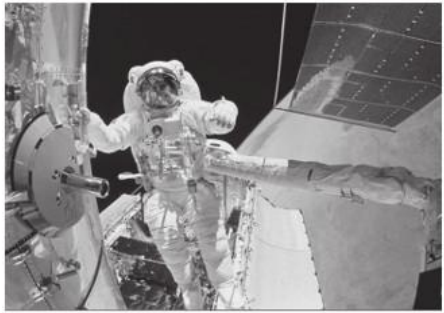

**Figura 2:** Questão 27 da prova azul de Ciências da Natureza do Enem 2009.

Na Tabela 3, estão apresentados os percentuais de escolha de cada uma das alternativas de resposta. A resposta considerada correta pelo Inep é a alternativa D, "pode-se afirmar que a frase dita pelo astronauta *não se justifica, porque a força peso é a força exercida pela gravidade terrestre, neste caso, sobre o telescópio e é a responsável por manter o próprio telescópio em órbita*", com um baixo percentual de acerto, de aproximadamente 14%.

**Tabela 3.** Percentual de escolha de cada uma das alternativas da questão (o gabarito é a letra D).

| 2009_Q27 | Percentual de acerto (%) |
|---|---|
| A | 42,1 |
| B | 20,5 |
| C | 6,5 |
| **D** | **13,9** |
| E | 16,7 |

As curvas características da questão 2009_Q27 são mostradas na Figura 3. A Figura 3a indica que a curva característica não se comporta de acordo com a previsão do modelo, segundo o qual a curva deve ser um S partindo de um valor reduzido e atingindo o valor 1 para altos escores. O ajuste está discrepante dos dados empíricos para altos escores. O parâmetro de acerto casual é da ordem de 11%, mas a discriminação do item é ligeiramente negativa. Uma discriminação negativa indica que estudantes com melhor desempenho (maior aptidão) possuem menor probabilidade de acerto da questão.

As curvas apresentadas na Figura 3b tem seus pontos determinados a partir de faixas obtidas de 20 em 20 pontos do escore (por exemplo, o ponto com nota 500 corresponde à faixa que vai de 490 a 510 pontos). Desta figura, verifica-se que o distrator (A), escolhido por 42% do total dos concluintes do ensino médio, é a mais marcada na faixa de escores de 260 a 760 pontos, com um pico em 660 pontos – quando é escolhida por 60% dos concluintes. Esta alternativa afirma que a frase dita pelo astronauta "*se* *justifica porque o tamanho do telescópio determina a sua massa, enquanto seu pequeno peso decorre da falta da ação da aceleração da gravidade."*

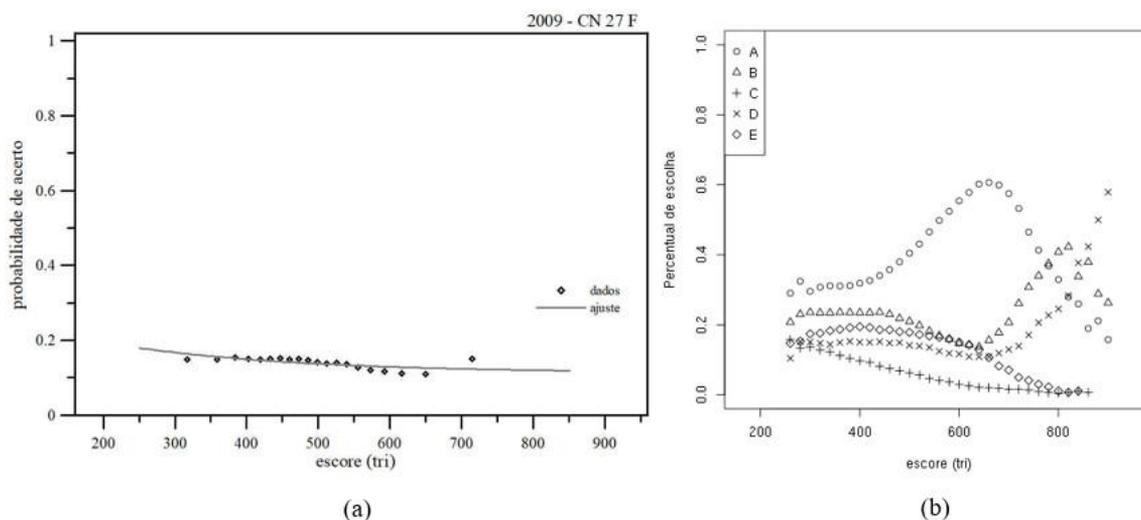

**Figura 3.** Questão 27 de 2009 (astronauta). (a) Curvas características do item – empírica (pontos) e ajuste pelo modelo logístico de 3 parâmetros (linha contínua). (b) Curvas que representam os percentuais de escolha de cada uma das alternativas em função da nota do aluno.

O distrator (B) apresenta um comportamento inesperado: parte de um percentual de escolha de cerca de 20% para baixos escores, decrescendo lentamente até o valor de cerca de 640 pontos, e a seguir aumenta rapidamente até um máximo no escore de 820 pontos. Esta alternativa, "*se justifica ao verificar que a inércia do telescópio é grande comparada à dele próprio, e que o peso do telescópio é pequeno porque a atração gravitacional criada por sua massa era pequena*", foi escolhida por 20% dos concluintes.

As alternativas (C) e (E) apresentam o comportamento esperado, decrescendo à medida que os escores dos estudantes aumentam.

É curioso observar que tanto o distrator (B) quanto o gabarito (D) possuem seus pontos de mínimo na mesma região em que o distrator (A) atinge seu máximo de escolhas. Ele é escolhido pela maioria dos concluintes, e afirma que o peso é pequeno como consequência da falta de ação da aceleração da gravidade. Em outras palavras, evidencia-se uma confusão conceitual entre peso e aceleração da gravidade, bem como a ideia de senso comum de que no espaço não há gravidade.

Portanto, a escolha do distrator (A) pela maioria dos estudantes indica a existência ao final da educação básica de uma concepção equivocada a respeito do conceito de gravidade, que associa a sua existência à existência de atmosfera [13;22;25]. Chegam ao final deste ciclo entendendo que é preciso ar como meio transmissor da força gravitacional. Corpos flutuam no espaço devido à inexistência de atmosfera e, consequentemente, de força gravitacional. O gabarito (D) só se torna mais marcado que os distratores nas últimas faixas de notas, acima de 800 pontos. Uma reflexão a ser feita é que a presença de um distrator tão atraente, como é o caso aqui, faz com que a curva característica não se comporte de acordo com a previsão do modelo.

**As dificuldades relativas à cinemática e a questão 2012_Q60**

O estudo das concepções sobre movimento data dos trabalhos de Piaget com crianças, sendo posteriormente estendido para vários níveis de ensino até o das disciplinas introdutórias de cursos de nível superior. A confusão conceitual entre posição e velocidade [27] e a dificuldade na construção e interpretação de gráficos [28] estão bem discutidas na literatura de pesquisa em ensino de física. O exemplo mais comum é a avaliação de que dois carros têm velocidades iguais ao se cruzarem quando se encontram numa mesma posição.

Na prova de Ciências da Natureza do Enem há alguns exemplos que podem evidenciar a presença desses equívocos. A questão 2012_Q60, apresentada na Figura 4, solicita que o estudante transforme informações dadas referentes a três instantes de tempo distintos para um trajeto de metrô em um gráfico posição versus tempo.

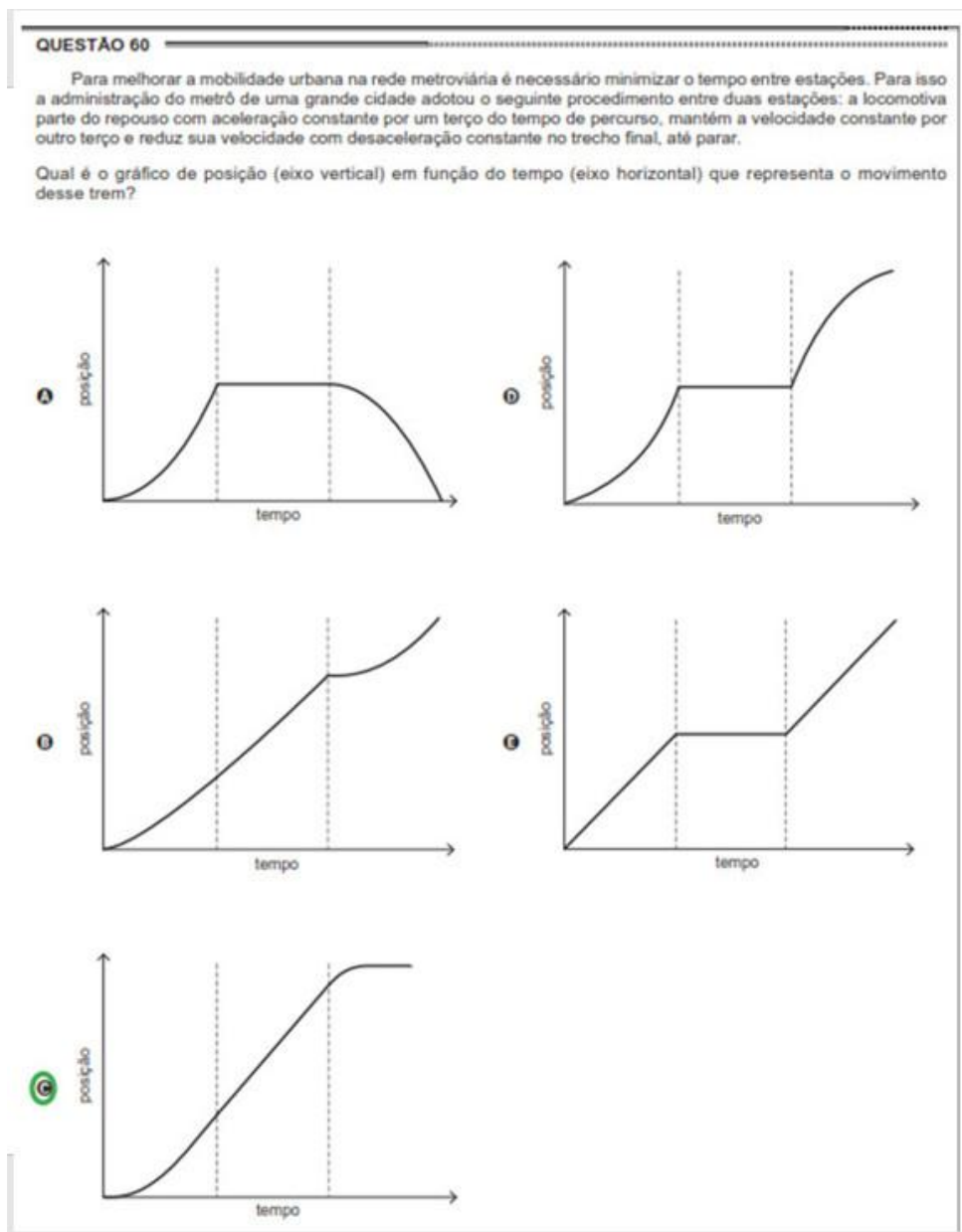

**Figura 4:** Questão 60 da prova azul de Ciências da Natureza do Enem 2012.

**Tabela 4.** Percentual de escolha de cada uma das alternativas da questão (o gabarito é a letra C).

| 2012_Q60 | Percentual de acerto (%) |
|---|---|
| A | 41,9 |
| B | 9,8 |
| **C** | **24,4** |
| D | 6,6 |
| E | 17,1 |

As alternativas (A), (D) e (E) são excluídas pela descrição do segundo intervalo de tempo, durante o qual o trem move-se com velocidade constante (não nula), e não mantém sua posição constante como previsto nessas três alternativas – que são a opção de 66% dos concluintes. O gabarito, (C), é escolhido por 24%.

As curvas características da questão são apresentadas na Figura 5a, e as curvas referentes ao percentual de escolha para cada uma das alternativas estão apresentadas na Figura 5b. A CCI é bem comportada, apresentando uma boa capacidade de discriminação e um parâmetro de dificuldade elevado, de 647 pontos (cerca de dois desvios padrão acima da média), com acerto casual moderado, em torno de 20%. As alternativas (A) e (C) são as duas mais assinaladas, como mostrado na Tabela 4, e juntas somam cerca de dois terços das respostas. As curvas traçadas nessas duas alternativas possuem o mesmo comportamento no primeiro trecho do movimento, correspondente a um movimento uniformemente acelerado, mas diferem nos dois outros trechos. No segundo trecho, a alternativa (A) indica que os concluintes podem ter confundido "velocidade constante" com "parado na mesma posição" e, no terceiro trecho, na alternativa (A), a informação do texto "até parar" não está correto. Os concluintes parecem confundir "passar pela origem" com "estar em repouso". O gráfico desta alternativa descreveria corretamente o movimento caso fosse um gráfico de velocidade (e não de posição) como função do tempo.

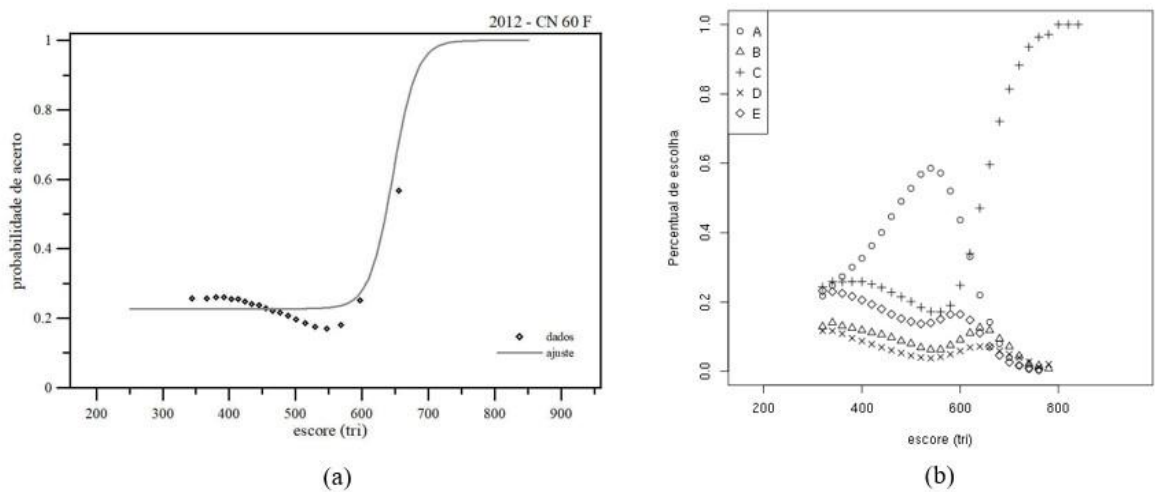

**Figura 5**. Questão 60 de 2012 (metrô). (a) Curvas características do item. (b) Percentuais de escolha de cada uma das alternativas em função da nota do aluno.

A Figura 5b nos informa que a alternativa (A) é a mais escolhida para respondentes com nota abaixo de cerca de 620 pontos. A partir deste valor, a opção correta (C) cresce monotonicamente. As demais alternativas revelam o comportamento esperado, diminuindo com o aumento da nota.

O desempenho de concluintes nesta questão deixa claro que ao final do ensino médio ainda há dificuldades em representar informações em forma de gráfico, e também de interpretar o que é apresentado de forma visual. A confusão entre os conceitos de posição e velocidade manifesta-se aqui de forma surpreendente, pois a questão que é aparentemente simples apresenta um parâmetro de dificuldade alto de 650 pontos, um e meio desvio padrão acima da média.

**As concepções sobre força e movimento e as questões 2013_Q87, 2014_Q67 e 2014_Q82**

Há muitas concepções não científicas relacionadas ao tema força e movimento; segundo Duit, Niederer e Schecker [29], "*a física é o campo no qual a maior parte das atividades de pesquisa nas concepções dos estudantes e em mudança conceitual foram desenvolvidas. (...) Torna-se óbvio que há uma ênfase particular em mecânica e eletricidade.*"[4] (p. 606). Sendo assim, podem-se destacar algumas relativas a este tema: força é proporcional à velocidade, se há movimento há força agindo sobre o corpo e esta força está na direção do movimento, peso e massa são a mesma coisa, se a velocidade é nula a aceleração também é.

O primeiro exemplo está na questão 87 da prova de Ciências da Natureza do Enem 2013, mostrada na Figura 6. O movimento descrito no enunciado possui duas partes, antes e depois do acionamento do paraquedas. No entanto, não há menção no texto a um sentido positivo ou negativo do eixo que descreve a vertical, e os gráficos apresentados nas alternativas parecem sugerir que o sentido positivo é o vertical para baixo (o sentido da força peso).

Da Figura 6, pode-se observar que as alternativas (A), (B) e (D) apresentam gráficos semelhantes na primeira parte do movimento, e o mesmo ocorre com as alternativas (C) e (E).

Na Tabela 5, apresentam-se os percentuais de escolha pelos concluintes de cada uma das alternativas. Há dois distratores, (A) e (D), com percentuais praticamente idênticos (25% e 26%, respectivamente), e a opção (C) possui 22% das marcações. O gabarito, (B), foi escolhido por 15% dos concluintes.

---

[4] Versão do original em inglês, "physics is the domain in which most research studies on investigating students' conceptions and on conceptual change have been carried out. (…) It becomes obvious that there is a pareticular emphasis on mechanics and eletricity."(p. 606)

**QUESTÃO 87**

Em um dia sem vento, ao saltar de um avião, um paraquedista cai verticalmente até atingir a velocidade limite. No instante em que o paraquedas é aberto (instante $T_A$), ocorre a diminuição de sua velocidade de queda. Algum tempo após a abertura do paraquedas, ele passa a ter velocidade de queda constante, que possibilita sua aterrissagem em segurança.

Que gráfico representa a força resultante sobre o paraquedista, durante o seu movimento de queda?

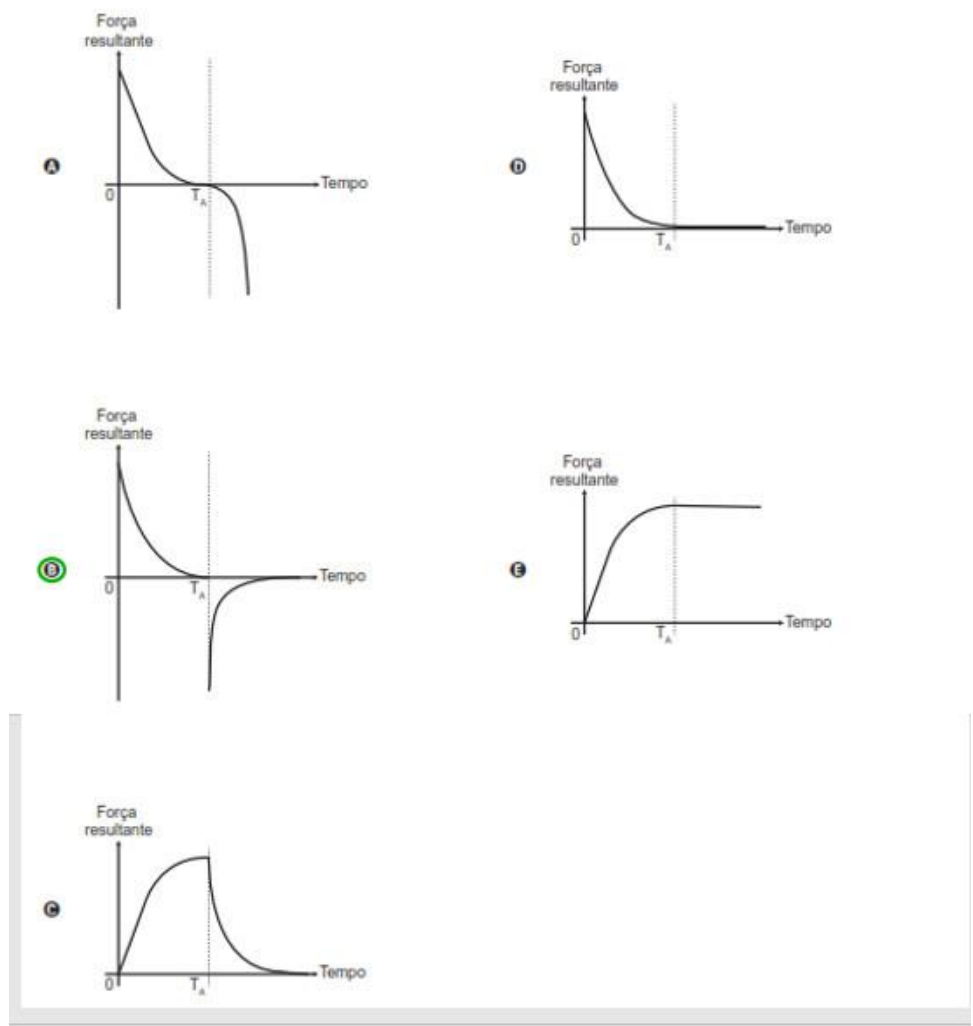

**Figura 6:** Questão 87 da prova azul de Ciências da Natureza do Enem 2013.

**Tabela 5.** Percentual de escolha de cada uma das alternativas da questão (o gabarito é a letra B).

| 2013_Q87 | Percentual de acertos (%) |
|---|---|
| A | 25,4 |
| **B** | **15,2** |
| C | 22,3 |
| D | 25,7 |
| E | 11,0 |

Nas Figuras 7a e 7b, estão as curvas características da questão e as curvas referentes ao percentual de escolha para cada uma das alternativas em função da nota. A CCI, na Figura 7a, correspondente ao ajuste do modelo apresenta um comportamento inadequado, com uma ligeira discriminação negativa (a frequência de acerto diminui quando a nota dos alunos aumenta), e os dados empíricos são um pouco diferentes do modelo para altos valores de notas.

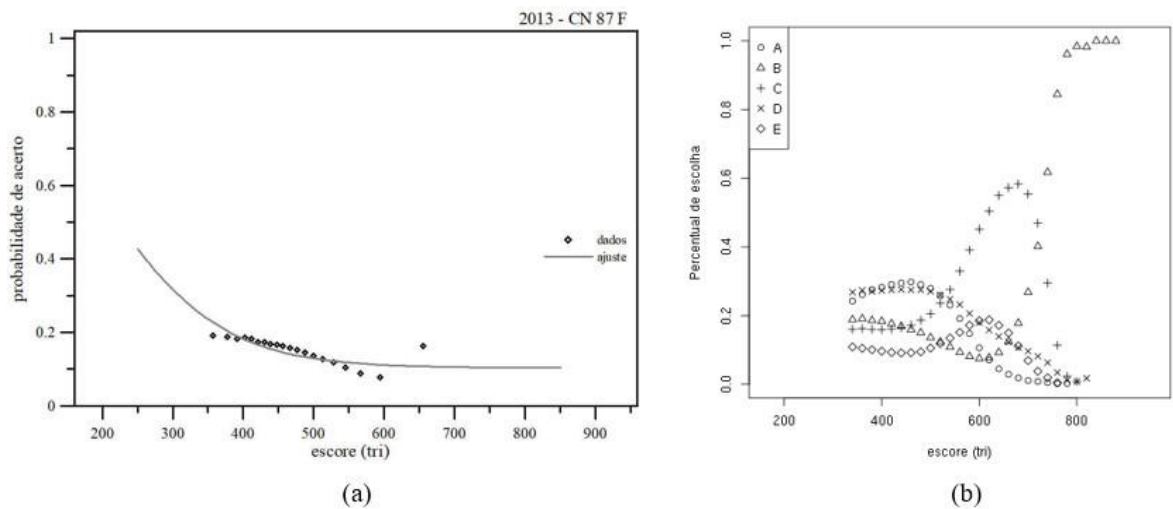

**Figura 7**. Questão 87 de 2013 (paraquedas). (a) Curvas características do item. (b) Percentuais de escolha de cada uma das alternativas em função da nota do aluno.

Na Figura 7b pode-se observar que os distratores (A) e (D) são as alternativas mais marcadas até cerca de 500 pontos, entre 540 e 720 pontos a opção (C) é a mais escolhida e, nesta região, também há um pico da opção (E). A escolha da alternativa (E) evidencia uma confusão conceitual da relação entre força e velocidade. Ou seja, relaciona-se com a concepção alternativa de que força é proporcional à velocidade [13]. Se a velocidade assume um valor constante, a força também é constante. É curioso notar que esta questão possui índice de discriminação negativo (a = -0,012). Isso pode ser visualizado pelo percentual de marcação da opção correta (B), que parte de aproximadamente 20% no limite inferior da escala de notas, cai abaixo de 10% por volta de 600 pontos quando começa a crescer e apenas a partir dos 740 pontos (para o grupo que tem nota cerca de 2,5 desvios padrão acima da média) torna-se a mais marcada.

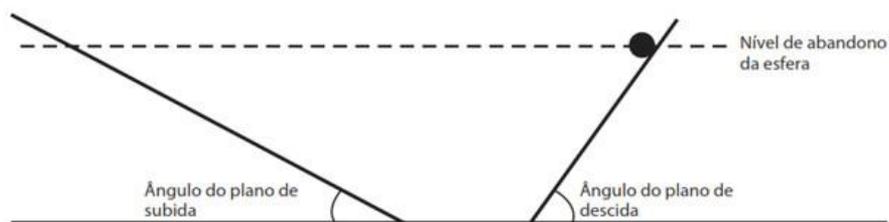

**Figura 8.** Questão 67 da prova azul de Ciências da Natureza do Enem 2014.

Outra questão que envolve este tipo de concepção é a questão 67 de 2014, mostrada na Figura 8. Nela há dois planos inclinados, e um deles torna-se horizontal. Pergunta-se o que acontece com uma esfera que desce um dos planos e continua o movimento na horizontal. Na Tabela 6, são apresentados os percentuais de escolha de cada uma das alternativas.

**Tabela 6.** Percentual de escolha de cada uma das alternativas da questão (o gabarito é a letra A).

| 2014_Q67 | Percentual de acertos (%) |
|---|---|
| **A** | **14,2** |
| B | 20,9 |
| C | 32,5 |
| D | 14,6 |
| E | 17,6 |

As curvas características da questão são mostradas na Figura 9a. Observa-se que se trata de uma questão com alta dificuldade (663 pontos, cerca de 2,5 desvios padrão acima da média), com alta capacidade de discriminação e baixo acerto casual (13%). Na Tabela 6, verifica-se que o distrator (C) é o que tem maior percentual de escolha, 33%. O fenômeno é descrito incorretamente no texto, e está presente a ideia não científica de que é necessária a presença de uma força agindo continuamente sobre um corpo para que ele permaneça em movimento, existindo uma relação linear entre força e velocidade [13,30].

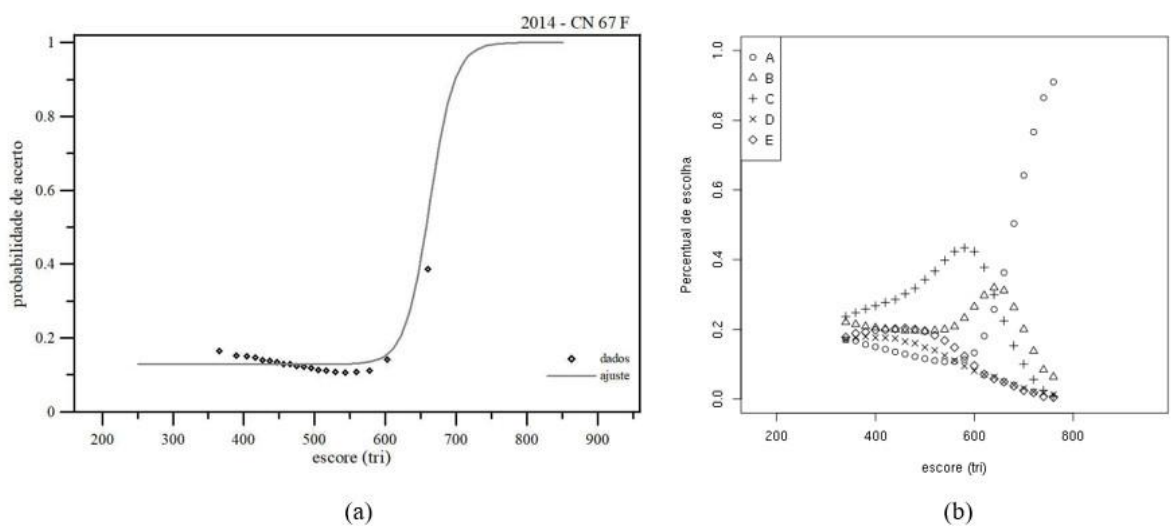

**Figura 9**. Questão 67 de 2014 (plano inclinado). (a) Curvas características do item. (b) Percentuais de escolha de cada uma das alternativas em função da nota do aluno.

Ao analisar a curva característica dos distratores em função da faixa de notas, na Figura 9b, observa-se que o distrator (C) é predominante, crescendo continuamente, até concluintes com nota de cerca de 580 pontos, quando começa a cair até ser superado pelas alternativas (B) e (A). O distrator (B) descreve corretamente o comportamento da velocidade, porém apresenta uma justificativa incorreta, a de que o impulso da descida continuaria a empurrá-la – uma concepção não científica de que se há movimento é porque há uma força agindo [13;30]. Esta alternativa apresenta um máximo de marcação percentual em cerca de 640 pontos. Somente a partir de 680 pontos o gabarito (A), que descreve o fenômeno e o explica adequadamente a partir do conceito de inércia, representa mais da metade das marcações.

Outro exemplo é apresentado na questão 82 da prova de 2014, na Figura 10. Trata-se da abordagem do movimento de satélites, usualmente descrito como circular uniforme cujas concepções também foram fruto de diversos trabalhos [30;31;32]. No caso, a pergunta é conceitual: espera-se que o estudante seja capaz de reconhecer as características dos vetores velocidade e aceleração no movimento circular uniforme.

Na Tabela 7, estão apresentados os percentuais de escolha de cada uma das alternativas. O gabarito (A) é assinalado por um número muito pequeno de concluintes, 12%. As alternativas (B) e (C), indicando que a aceleração tangencial tem a mesma direção da velocidade, são escolhidas por 48% dos concluintes; as alternativas (D) e (E), para as quais a aceleração tangencial é perpendicular à velocidade, são escolhidas por 39% dos respondentes.

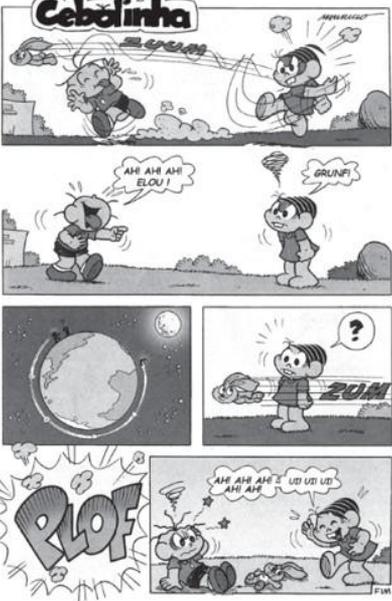

**Figura 10:** Questão 82 da prova azul de Ciências da Natureza do Enem 2014.

**Tabela 7.** Percentual de escolha de cada uma das alternativas da questão (o gabarito é a letra A).

| 2014_Q82 | Percentual de acerto (%) |
|---|---|
| **A** | **12,2** |
| B | 35,1 |
| C | 13,1 |
| D | 19,1 |
| E | 20,3 |

Na Figura 11a, estão as curvas características da questão, e na Figura 11b as curvas referentes ao percentual de escolha para cada uma das alternativas em função da nota. A CCI é bem comportada, revelando uma questão extremamente difícil, com parâmetro de dificuldade de aproximadamente 700 pontos, que discrimina muito os estudantes a partir de cerca de 650 pontos, e que possui um baixo parâmetro de acerto casual. O distrator mais marcado, com 35% das escolhas, é a opção (B), alternativa para a qual o movimento circular não é uniforme, indicando um desconhecimento do concluinte quanto ao papel das componentes da aceleração no movimento curvilíneo. Talvez, porém, possa-se inferir que esta marcação revela a presença de uma concepção equivocada de que se há movimento, há uma força na direção do movimento e que esta força é proporcional à velocidade, reforçando a ideia do "impetus". Este distrator é a principal escolha de candidatos com nota inferior a 600 pontos, crescendo até cerca de 480 pontos e depois diminuindo; o distrator (D) apresenta o mesmo tipo de comportamento, crescendo para notas até cerca de 620 pontos; é a escolha dominante dos alunos entre os valores de 600 e 700 pontos (já acima da média das notas). Já a escolha do distrator (E) evidencia a crença de que sobre o coelho há uma força no movimento circular dirigida para fora do círculo, a força centrífuga. Somente a partir dos 700 pontos é que o percentual de escolha da opção correta (A) passa a prevalecer.

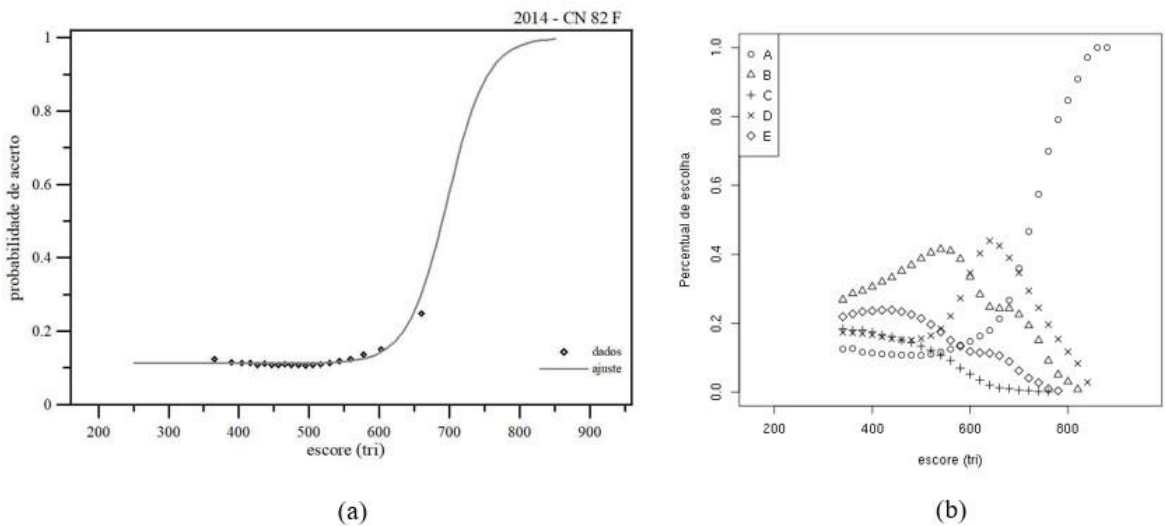

**Figura 11**. Questão 82 de 2014 (Mônica). (a) Curvas características do item. (b) Percentuais de escolha de cada uma das alternativas em função da nota do aluno.

O desempenho dos estudantes concluintes no ensino médio nessas questões indica que as dificuldades conceituais conhecidas no tema da relação entre força e movimento, continuam muito presentes e impactando na aprendizagem dos estudantes ao final do ensino médio.

**Flutuação dos corpos e a questão 2010_Q81**

A questão 2010_Q81, mostrada na Figura 12, discute o preenchimento de uma piscina com água para auxiliar na remoção de uma escultura e solicita que o estudante escolha a opção que explica o porquê da remoção ficar mais fácil.

**Questão 81**

Durante uma obra em um clube, um grupo de trabalhadores teve de remover uma escultura de ferro maciço colocada no fundo de uma piscina vazia. Cinco trabalhadores amarraram cordas à escultura e tentaram puxá-la para cima, sem sucesso.

Se a piscina for preenchida com água, ficará mais fácil para os trabalhadores removerem a escultura, pois a

Se a piscina for preenchida com água, ficará mais fácil para os trabalhadores removerem a escultura, pois a

Ⓐ escultura flutuará. Dessa forma, os homens não precisarão fazer força para remover a escultura do fundo.
Ⓑ escultura ficará com peso menor. Dessa forma, a intensidade da força necessária para elevar a escultura será menor.
Ⓒ água exercerá uma força na escultura proporcional a sua massa, e para cima. Esta força se somará à força que os trabalhadores fazem para anular a ação da força peso da escultura.
Ⓓ água exercerá uma força na escultura para baixo, e esta passará a receber uma força ascendente do piso da piscina. Esta força ajudará a anular a ação da força peso na escultura.
Ⓔ água exercerá uma força na escultura proporcional ao seu volume, e para cima. Esta força se somará à força que os trabalhadores fazem, podendo resultar em uma força ascendente maior que o peso da escultura.

**Figura 12.** Questão 81 da prova azul de Ciências da Natureza do Enem 2010.

Na Tabela 8, estão apresentados o percentual de escolha de cada uma das alternativas apresentadas.

**Tabela 8.** Percentual de escolha de cada uma das alternativas da questão (o gabarito é a letra E).

| 2010_q81 | Percentual de acertos (%) |
|---|---|
| A | 7,8 |
| B | 25,4 |
| C | 22,4 |
| D | 12,3 |
| **E** | **31,6** |

Na Figura 13, estão as curvas características da questão. A alternativa mais escolhida é o gabarito, porém o percentual de acerto não é alto (32%). As curvas características da Figura 13a indicam que esta é uma questão difícil, com parâmetro de dificuldade de cerca de 650 pontos, com discriminação adequada. O percentual de acerto casual é cerca de 22%.

As alternativas (B) e (C) também são muito marcadas, com 25% e 22% das escolhas, respectivamente. Os concluintes indicam saber que mesmo com a piscina cheia a escultura não flutuará, já que apenas 8% deles assinalam a resposta (A).

Na opção (B), é feita confusão entre a redução da força necessária para levantar a escultura e a redução de seu peso, que só poderia ocorrer com a diminuição da massa da escultura ou com a redução da aceleração da gravidade. Esta é uma concepção não científica muito comum quando se estuda a flutuação de corpos [33,34,35]. Na Figura 13b, observa-se que (B) é a alternativa mais marcada até cerca de 500 pontos, quando é ultrapassada pela opção correta (E). Já o distrator (C) afirma erroneamente que as forças exercidas pela água (empuxo) e pelos trabalhadores (tensão na corda) sobre a escultura anulam a força peso que age sobre ela. A marcação desta alternativa possui um máximo de quase 30% em 600 pontos e ainda se mantém acima de 10% até os 700 pontos. Em outras palavras, as concepções evidenciadas nos distratores (B) e (C) estão fortemente presentes em quase metade dos respondentes.

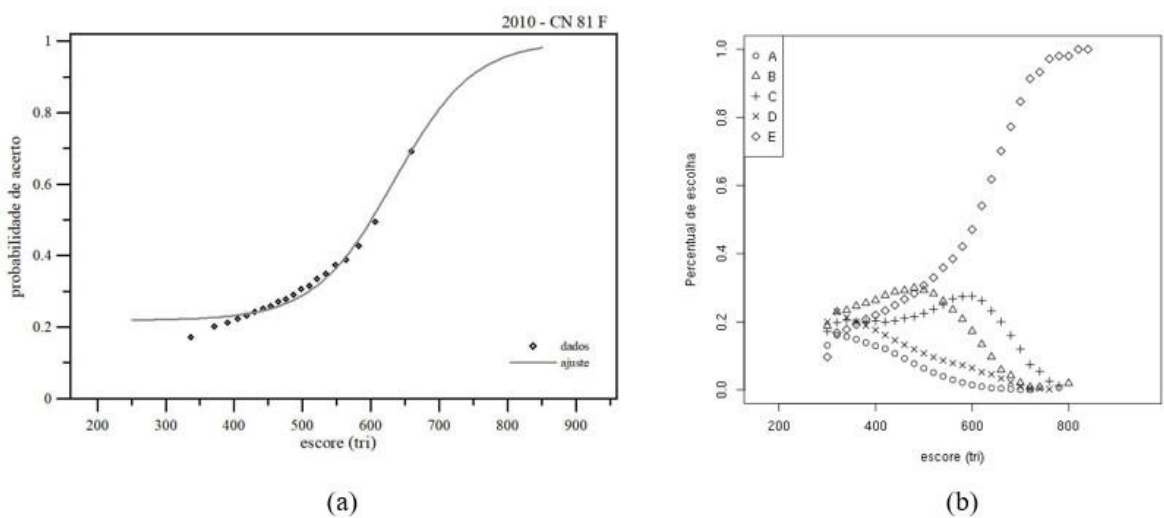

**Figura 13**. Questão 81 de 2010 (escultura). (a) Curvas características do item. (b) Percentuais de escolha de cada uma das alternativas em função da nota do aluno.

Esta questão demonstra que os estudantes chegam ao final do ensino médio com deficiências conceituais básicas tanto no entendimento das características e da natureza da força peso quanto na identificação das forças que agem na escultura.

**As concepções sobre atrito, a questão 2013_Q76 e a mecânica universitária básica**

Outro nó conceitual refere-se à compreensão da força de atrito. Caldas e Saltiel [36] discutem que a abordagem de forma "negativa" usualmente adotada na discussão sobre a força de atrito estática ou cinética em escolas ou disciplinas introdutórias de graduação na França, normalmente apresentada como uma força de resistência ou dissipativa, ou num contexto de frenagem, leva à concepção de que esta força é por definição contrária ao movimento.

A questão 2013_Q76, mostrada na Figura 14, revela que no Brasil a situação não é diferente. Esta questão aborda justamente quais devem ser a direção e o sentido da força de atrito nos pés de uma pessoa que se desloca sobre uma superfície. Na Tabela 9, estão indicados os percentuais de escolha de cada uma das alternativas. Nelas, pode-se escolher para a direção da força de atrito nos pés a direção perpendicular ao plano, (A), escolhida por 17% dos respondentes; a direção paralela ao plano, (B) e (C), escolhida por 50% dos respondentes; a direção horizontal, (D), com 14% e a direção vertical, (E) com 19% das escolhas. A escolha da direção paralela ao plano permite a opção pelo sentido: contrário ao movimento (30%) e no mesmo sentido do movimento (20%).

## QUESTÃO 76

Uma pessoa necessita da força de atrito em seus pés para se deslocar sobre uma superfície. Logo, uma pessoa que sobe uma rampa em linha reta será auxiliada pela força de atrito exercida pelo chão em seus pés.

Em relação ao movimento dessa pessoa, quais são a direção e o sentido da força de atrito mencionada no texto?

Ⓐ Perpendicular ao plano e no mesmo sentido do movimento.
Ⓑ Paralelo ao plano e no sentido contrário ao movimento.
Ⓒ Paralelo ao plano e no mesmo sentido do movimento.
Ⓓ Horizontal e no mesmo sentido do movimento.
Ⓔ Vertical e sentido para cima.

**Figura 14.** Questão 76 da prova azul de Ciências da Natureza do Enem 2013.

**Tabela 9.** Percentual de escolha de cada uma das alternativas da questão (o gabarito é a letra C).

| 2010_q81 | Percentual de acertos (%) |
|---|---|
| A | 16,7 |
| B | 30,1 |
| **C** | **19,9** |
| D | 14,3 |
| E | 18,7 |

A curva característica da questão é mostrada na Figura 15a. O ajuste do modelo apresenta discriminação negativa, indicando que estudantes com notas menores acertam mais do que os com notas maiores. E isso pode ser melhor entendido a partir dos gráficos de percentual de escolha dos distratores em função da nota, mostradas na Figura 15b. Todas as alternativas são escolhidas por cerca de 20% dos respondentes com nota por volta de 440 pontos. A partir desta nota (em torno de meio desvio padrão abaixo da média, e aproximadamente a nota necessária para obtenção da certificação do ensino médio na época), os distratores (A), (D) e (E) perdem a atratividade para os respondentes. E a opção (B), corresponde à concepção persistente e reforçada em materiais didáticos [37] de que o atrito sempre se opõe ao movimento, passa a dominar as escolhas, obtendo 70% das respostas para respondentes com 640 pontos. O gabarito (C) só passa a ser a alternativa mais marcada acima de 720 pontos. Mesmo entre candidatos com notas altas (com mais de dois desvios padrão acima da média) existe uma parcela significativa de estudantes que possui um conhecimento limitado e incorreto sobre o comportamento da força de atrito.

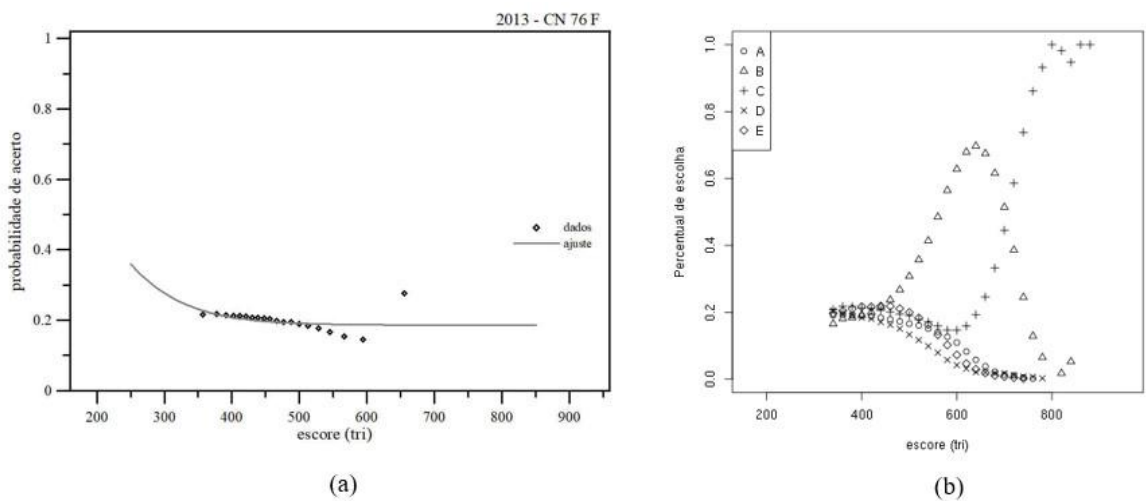

**Figura 15.** Questão 76 de 2013 (atrito). (a) Curvas características do item. (b) Percentuais de escolha de cada uma das alternativas em função da nota do aluno.

Esta persistência de uma concepção sobre o comportamento da força de atrito foi explorada, a partir da observação dos resultados do Enem, em duas situações: na primeira, como descrito por Rinaldi [12], foi apresentada a questão a um grande número de estudantes do ensino médio em uma escola da rede pública federal do estado do Rio de Janeiro, com a solicitação de justificar por escrito a escolha – e o resultado obtido confirmou os achados aqui apresentados. Na segunda, a questão (em formato similar, também de múltipla escolha) foi apresentada em uma prova da disciplina de Mecânica Introdutória (Física I) na UFRJ. A questão está apresentada na Figura 16. Na Tabela 10, apresenta-se o percentual de escolha de cada uma das alternativas[5].

5. Uma pessoa caminha, subindo um plano inclinado, com velocidade constante. Só atuam sobre a pessoa as forças peso, normal e a de atrito que o plano inclinado exerce sobre ela. É correto afirmar que a força de atrito que o plano exerce sobre o pé da pessoa:

(a) É paralela ao plano, no sentido da velocidade da pessoa.
(b) É paralela ao plano, no sentido contrário ao da velocidade da pessoa.
(c) É perpendicular ao plano, no sentido da força normal.
(d) É perpendicular ao plano, no sentido contrário ao da força normal.
(e) Tem a mesma direção mas sentido contrário ao da força peso.

**Figura 16.** Questão apresentada a 1200 estudantes no ano inicial de um curso universitário.

**Tabela 10.** Percentual de escolha de cada uma das alternativas da questão (o gabarito é a letra A).

| Alternativa | Percentual de escolha |
|---|---|
| **A** | **56.7** |
| B | 40.9 |
| C | 0.3 |
| D | 1.3 |
| E | 0.6 |

Observa-se da Tabela 10 que as escolhas de estudantes universitários se concentram em respostas que indicam que a força de atrito está na direção do plano. No entanto, mesmo com a dominância das respostas de que o atrito tem a direção do movimento, ainda há um número significativo (41%) de estudantes que mantêm a concepção de que o atrito é sempre oposto ao movimento. Na Figura 17, está apresentada a curva característica desta questão (a nota do aluno é apresentada em unidades estatísticas, com 0 sendo a média e 1 o desvio padrão). Observa-se que mesmo com muitos alunos escolhendo a alternativa em que o atrito é oposto ao movimento, a curva característica tem o comportamento esperado – à medida que a nota do estudante aumenta, o percentual de acerto também aumenta.

---

[5] Fonte: M.F. Barroso, comunicação privada.

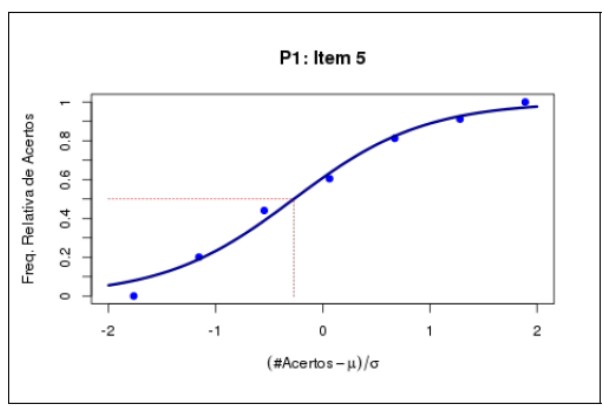

**Figura 17.** Curva característica do item (modelo logístico de dois parâmetros e dados empíricos) para a questão mostrada na Figura 16.

O resultado evidencia que uma parcela significativa de estudantes no final do ensino médio ou no início de um curso universitário desenvolvem uma concepção equivocada de que a força de atrito é, por definição, contrária ao movimento. Essa concepção parece ter origem em uma abordagem "negativa" adotada no processo de instrução e reforçada por materiais didáticos.

**A relação entre força e pressão e as questões 2012_Q47 e 2013_Q61**

A dificuldade de compreensão do conceito de pressão e a sua diferenciação do conceito de força [33;38] pode ser estudado a partir da observação do desempenho dos concluintes nas questões 2012_Q47 e 2013_Q61.

A discussão na questão 47 da prova de 2012, mostrada na Figura 18, é conceitual. O percentual de escolha de cada uma das alternativas está indicado na Tabela 11.

**QUESTÃO 47**

Um dos problemas ambientais vivenciados pela agricultura hoje em dia é a compactação do solo, devida ao intenso tráfego de máquinas cada vez mais pesadas, reduzindo a produtividade das culturas.

Uma das formas de prevenir o problema de compactação do solo é substituir os pneus dos tratores por pneus mais

Ⓐ largos, reduzindo a pressão sobre o solo.
Ⓑ estreitos, reduzindo a pressão sobre o solo.
Ⓒ largos, aumentando a pressão sobre o solo.
Ⓓ estreitos, aumentando a pressão sobre o solo.
Ⓔ altos, reduzindo a pressão sobre o solo.

**Figura 18.** Questão 47 da prova azul de Ciências da Natureza do Enem 2012.

**Tabela 11.** Percentual de escolha de cada uma das alternativas da questão (o gabarito é a letra A).

| 2012_q47 | Percentual de acerto (%) |
|---|---|
| **A** | **43,9** |
| B | 32,6 |
| C | 3,8 |
| D | 2,6 |
| E | 16,9 |

As curvas características dessa questão são apresentadas na Figura 19a. Trata-se de uma questão de dificuldade moderada, cerca de 540 pontos, com boa capacidade de discriminação e parâmetro de acerto casual correspondente a 23%.

As alternativas mais marcadas, o gabarito (A) com 44% das escolhas, (B) com 33% e (E) com 17%, de acordo com a Tabela 11, indicam que a maioria dos concluintes sabe que a pressão deve ser reduzida; no entanto, eles não sabem relacionar a característica do pneu que possibilita essa diminuição, se largo, estreito ou alto.

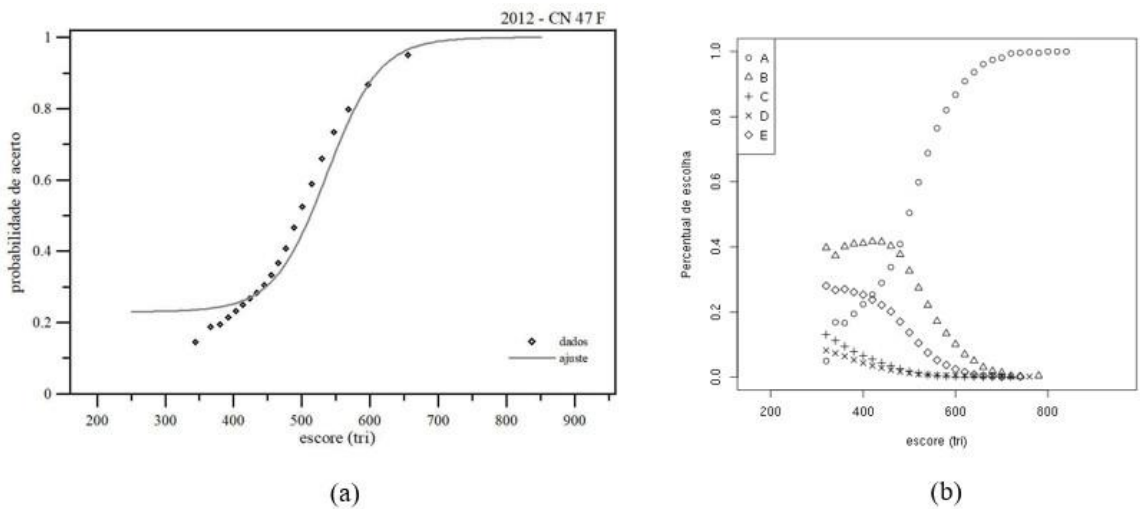

**Figura 19**. Questão 47 de 2012 (pneu). (a) Curvas características do item. (b) Percentuais de escolha de cada uma das alternativas em função da nota do aluno.

Apesar destas três opções mencionarem uma redução na pressão, as curvas características desses distratores são distintas, como mostrado na Figura 19b. O gabarito (A) ultrapassa os 50% de escolha somente a partir dos 500 pontos, com menos de 20% de marcação nas notas mais baixas, e somente ultrapassa as opções (E) e (B) a partir dos 420 e 480 pontos, respectivamente, indicando uma presença (para os respondentes com nota mais baixa) de dificuldades em distinguir os conceitos de força e pressão.

Já a questão 2013_Q61, cujo enunciado está mostrado na Figura 20, é de caráter quantitativo. Não há nenhuma imagem na apresentação da questão, o que exige que o concluinte imagine a situação proposta (e portanto devendo conhecer o elevador hidráulico). O gabarito é a alternativa (C), com 28% das escolhas; é necessário calcular o peso a ser sustentado pelo pistão, incluindo as massas da pessoa, da cadeira de roda e da plataforma, e lembrar que a pressão nos dois extremos é a mesma, mas as forças de sustentação não são as mesmas. A resposta (D), a mais marcada, com 33% das escolhas, corresponde a tomar como valores idênticos as forças dos dois lados do elevador hidráulico, confundindo, portanto, força com pressão. A alternativa (B), escolhida por 22% dos alunos, equivale numericamente a responder que a força é o valor da massa total.

As curvas características estão apresentadas na Figura 21a. O item apresenta uma dificuldade de 650 pontos, alta discriminação e alto acerto casual (em torno de 27%).

**QUESTÃO 61**

Para oferecer acessibilidade aos portadores de dificuldades de locomoção, é utilizado, em ônibus e automóveis, o elevador hidráulico. Nesse dispositivo é usada uma bomba elétrica, para forçar um fluido a passar de uma tubulação estreita para outra mais larga, e dessa forma acionar um pistão que movimenta a plataforma. Considere um elevador hidráulico cuja área da cabeça do pistão seja cinco vezes maior do que a área da tubulação que sai da bomba. Desprezando o atrito e considerando uma aceleração gravitacional de 10 m/s², deseja-se elevar uma pessoa de 65 kg em uma cadeira de rodas de 15 kg sobre a plataforma de 20 kg.

Qual deve ser a força exercida pelo motor da bomba sobre o fluido, para que o cadeirante seja elevado com velocidade constante?

Ⓐ 20 N
Ⓑ 100 N
Ⓒ 200 N
Ⓓ 1 000 N
Ⓔ 5 000 N

**Figura 20.** Questão 61 da prova azul de Ciências da Natureza do Enem 2013.

**Tabela 12.** Percentual de escolha de cada uma das alternativas da questão (o gabarito é a letra C).

| 2013_q61 | Percentual de acerto (%) |
|---|---|
| A | 7,9 |
| B | 22,3 |
| **C** | **27,6** |
| D | 33,2 |
| E | 8,7 |

Os percentuais de escolha das alternativas por faixa de nota são exibidos na Figura 21b. É perceptível que as opções mais marcadas possuem comportamentos distintos; as três alternativas, (B), (C) e (D) são igualmente escolhidas no limite de baixas notas, por cerca de 25% dos concluintes, mas a opção (B), cujo valor é a soma das massas colocadas sobre o pistão, cai rapidamente. A alternativa (D), correspondendo ao peso colocado sobre o pistão, é crescente entre os concluintes nas faixas de 400 a 600 pontos, sendo a mais escolhida até cerca de 640 pontos. Concluintes com nota inferior a 640 pontos, portanto, não entendem a diferença entre força e pressão, ignorando a informação "cuja área da cabeça do pistão seja cinco vezes maior do que a área da tubulação que sai da bomba" contida no texto e supondo que a força, e não a pressão, é igual dos dois lados do elevador. A partir de 640 pontos, o gabarito (C) passa a ter o maior percentual de escolha, indicando que a partir desta região o candidato consegue utilizar a informação da relação entre as áreas dos pistões de forma adequada. A opção (E) mantém-se abaixo de 10% de marcação com exceção da região entre 580 e 700 pontos, onde apresenta um pequeno pico que, provavelmente, representa um erro conceitual no qual utiliza que pressão é força multiplicada pela área.

Esta questão é quantitativa e requer que o estudante aplique o Princípio de Pascal, além de ter que imaginar o que é um elevador hidráulico já que não há imagem na questão. O percentual de acerto não é alto e a escolha do distrator mais atraente (D) evidencia o desconhecimento da diferença entre força e pressão.

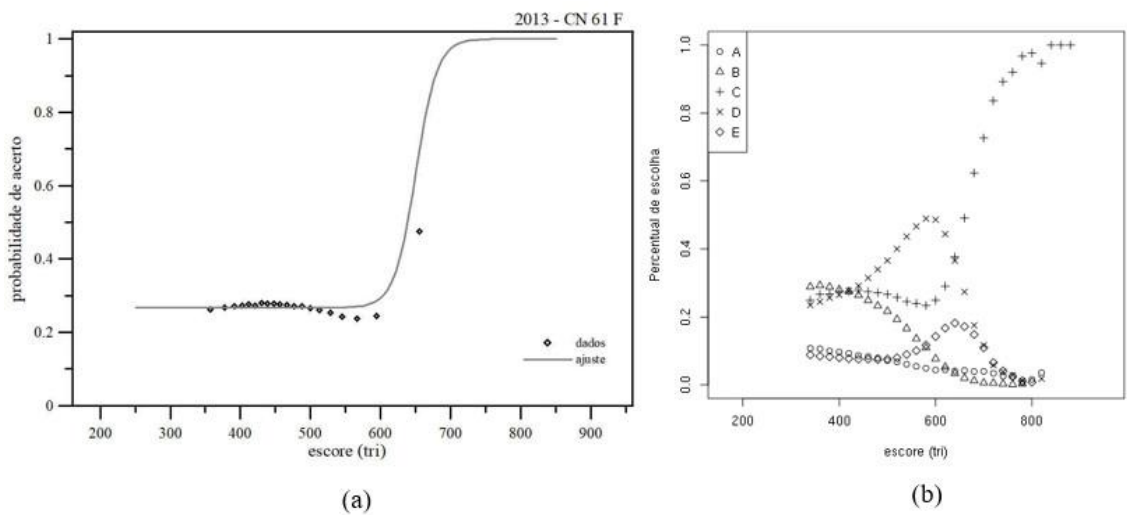

**Figura 21**. Questão 61 de 2013 (elevador hidráulico). (a) Curvas características do item. (b) Percentuais de escolha de cada uma das alternativas em função da nota do aluno.

### As concepções sobre temperatura e calor e a questão 2010_Q50

A percepção cotidiana relativa às sensações de "quente" e de "frio" não conduz a uma compreensão correta dos conceitos de temperatura e calor. Ao contrário, desenvolve-se uma noção intuitiva equivocada destes dois conceitos, e este é um dos temas mais abordados nos trabalhos sobre concepções não científicas dos estudantes [29]. Esse é o tema da questão 50 da prova do Enem 2010, na Figura 22. A pergunta feita, "Do ponto de vista científico, que situação prática mostra a limitação dos conceitos corriqueiros de calor e temperatura?" exige do concluinte não apenas a compreensão correta desses conceitos como também o contraste, na análise de uma afirmação relativa a um fenômeno do dia a dia, entre os pontos de vista cientificamente aceito e o do senso comum, um exercício cognitivamente difícil.

**Questão 50**

Em nosso cotidiano, utilizamos as palavras "calor" e "temperatura" de forma diferente de como elas são usadas no meio científico. Na linguagem corrente, calor é identificado como "algo quente" e temperatura mede a "quantidade de calor de um corpo". Esses significados, no entanto, não conseguem explicar diversas situações que podem ser verificadas na prática.

Do ponto de vista científico, que situação prática mostra a limitação dos conceitos corriqueiros de calor e temperatura?

Ⓐ A temperatura da água pode ficar constante durante o tempo em que estiver fervendo.
Ⓑ Uma mãe coloca a mão na água da banheira do bebê para verificar a temperatura da água.
Ⓒ A chama de um fogão pode ser usada para aumentar a temperatura da água em uma panela.
Ⓓ A água quente que está em uma caneca é passada para outra caneca a fim de diminuir sua temperatura.
Ⓔ Um forno pode fornecer calor para uma vasilha de água que está em seu interior com menor temperatura do que a dele.

**Figura 22.** Questão 50 da prova azul de Ciências da Natureza do Enem 2010.

**Tabela 13.** Percentual de escolha de cada uma das alternativas da questão (o gabarito é a letra A).

| 2010_q50 | Percentual de acerto (%) |
|---|---|
| **A** | **18,3** |
| B | 21,0 |
| C | 17,1 |
| D | 15,7 |
| E | 27,6 |

As ideias que representam a "resistência" da concepção cotidiana [39] são: calor é uma substância que sentimos pelo tato, ideia presente na alternativa (B) com 21% das escolhas; calor é uma substância que esquenta as coisas, presente nos distratores (C) e (E) com 17% e 28% das escolhas, respectivamente; temperatura mede a quantidade de calor, presente na opção (D), com 16% das escolhas.

A formulação da questão exige do respondente um raciocínio cognitivo de alto nível, pois é preciso vincular um fenômeno a uma explicação não correta. É necessária a reflexão sobre que tipo de fenômeno esses conceitos cotidianos não permitem explicar: a absorção de calor, a propagação de calor, que a sensação de tato mede fluxo de calor, e, entre outros, que se um corpo recebe calor nem sempre sua temperatura fica aumentada. É essa última reflexão a apresentada no gabarito (A) – escolhida apenas por 18% dos concluintes.

As curvas características apresentadas na Figura 23a mostram que a questão é extremamente difícil, com parâmetro de dificuldade 713. De acordo com o modelo, a discriminação é adequada e o acerto casual é de 17%. Da Figura 23b, observa-se que todas as opções são escolhidas com praticamente a mesma frequência para respondentes com notas até 450 pontos, e que a partir de 450 até cerca de 620 pontos a escolha mais frequente é o distrator (E) – alternativa que apresenta uma descrição correta de um fenômeno, mas que não responde à pergunta feita. A partir de 620 pontos, há dominância da escolha do gabarito (A).

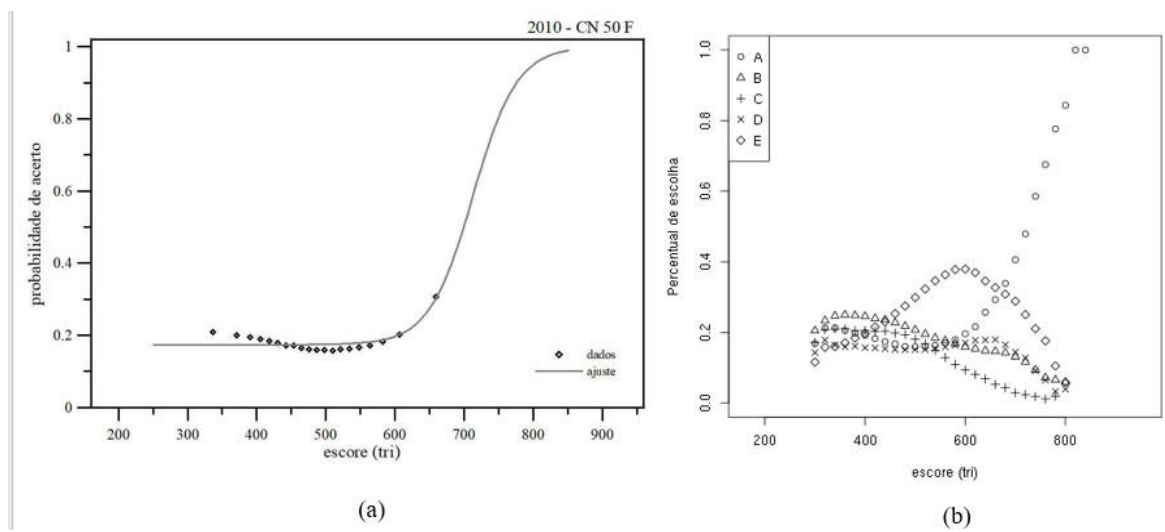

**Figura 23**. Questão 50 de 2010 (calor e temperatura). (a) Curvas características do item. (b) Percentuais de escolha de cada uma das alternativas em função da nota do aluno.

Outra concepção relativa às propriedades térmicas dos materiais é a de que objetos que prontamente se aquecem não se resfriam da mesma forma [40,41]. Na questão 48 de 2013, apresentada na Figura 24 e discutida em Rinaldi [12], essa concepção fica evidente. A afirmação "maior no aquecimento" está presente em três opções, (B), (D) e (E), somando 80% das escolhas. No entanto, o gabarito (E) é escolhido apenas por 18% dos respondentes, e o distrator (D), que evidencia a diferença entre aquecimento e resfriamento, tem 44% das escolhas.

## QUESTÃO 48

Em um experimento foram utilizadas duas garrafas PET, uma pintada de branco e a outra de preto, acopladas cada uma a um termômetro. No ponto médio da distância entre as garrafas, foi mantida acesa, durante alguns minutos, uma lâmpada incandescente. Em seguida a lâmpada foi desligada. Durante o experimento, foram monitoradas as temperaturas das garrafas: a) enquanto a lâmpada permaneceu acesa e b) após a lâmpada ser desligada e atingirem equilíbrio térmico com o ambiente.

A taxa de variação da temperatura da garrafa preta, em comparação à da branca, durante todo experimento, foi

Ⓐ igual no aquecimento e igual no resfriamento.
Ⓑ maior no aquecimento e igual no resfriamento.
Ⓒ menor no aquecimento e igual no resfriamento.
Ⓓ maior no aquecimento e menor no resfriamento.
Ⓔ maior no aquecimento e maior no resfriamento.

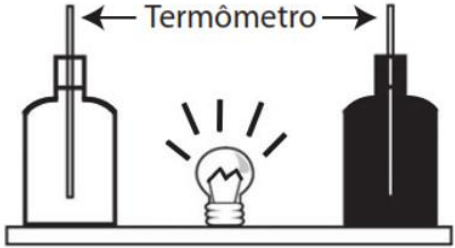

**Figura 21.** Questão 48 da prova azul de Ciências da Natureza do Enem 2013.

**Tabela 14.** Percentual de escolha de cada uma das alternativas da questão (o gabarito é a letra E).

| 2013_q48 | Percentual de acerto (%) |
|---|---|
| A | 10,3 |
| B | 18,2 |
| C | 9,5 |
| D | 43,5 |
| **E** | **18,3** |

Na Figura 25a, estão as curvas características. O parâmetro de dificuldade na questão é elevado, de 690 pontos, o acerto casual é de 13%, e a discriminação é adequada – mas o modelo não descreve com precisão os dados empíricos.

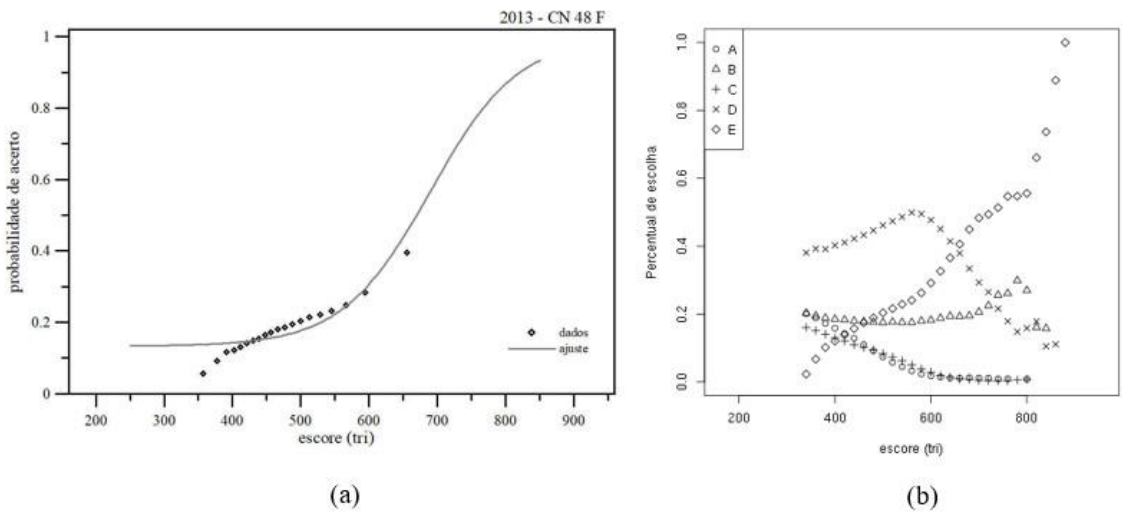

**Figura 25.** Questão 48 de 2013 (garrafa PET). (a) Curvas características do item. (b) Percentuais de escolha de cada uma das alternativas em função da nota do aluno.

Das curvas que apresentam os percentuais de escolha das alternativas por faixa de nota, na Figura 25b, observa-se que o distrator preferido pelos concluintes tem um máximo em 580 pontos e domina as respostas até 620 pontos, quando é preterida pelo gabarito. A alternativa (D) tem um comportamento interessante, crescendo o percentual de alunos que a escolhe até o valor de 800 pontos, quando é escolhida por 30% dos alunos – esta é a região de notas que está além de 3 desvios padrão acima da média.

O desempenho indicado por esses resultados evidencia claramente que as concepções não científicas relacionadas a calor e temperatura apresentam-se muito relevantes ao final do ensino médio.

**Os efeitos ambientais da geração de energia e a questão 2011_Q80**

A questão 2011_Q80, apresentada na Figura 26 e discutida em Lopes [11], exige somente a leitura atenta do texto para responder. Na Tabela 15, está indicado o percentual de escolha de cada uma das alternativas.

Apenas uma das alternativas aponta a fonte hidrelétrica como sendo poluidora, como indica o texto. Um estudo[6] realizado por pesquisadores da COPPE/UFRJ e apresentado na Conferência Rio 2002 mostra que barragens de hidrelétricas produzem quantidades consideráveis de metano, gás carbônico e óxido nitroso, gases que provocam o chamado efeito estufa. Esta alternativa, (D), é o gabarito, e foi escolhido por 21% dos concluintes.

O distrator (C) é o mais marcado, com 23% das escolhas, e baseia-se na ideia do senso comum que a energia proveniente de fontes hidrelétricas é limpa. Mas há uma distribuição quase equivalente entre todos os distratores, que apontam para essa forma de gerar energia como correspondendo a uma geração limpa. Ou seja, há uma visão equivocada desta forma de gerar energia [42,43], pois ela também é fonte de emissão de gases poluentes na atmosfera contribuindo para o crescimento do efeito estufa.

**QUESTÃO 80**

Segundo dados do Balanço Energético Nacional de 2008, do Ministério das Minas e Energia, a matriz energética brasileira é composta por hidrelétrica (80%), termelétrica (19,9%) e eólica (0,1%). Nas termelétricas, esse percentual é dividido conforme o combustível usado, sendo: gás natural (6,6%), biomassa (5,3%), derivados de petróleo (3,3%), energia nuclear (3,1%) e carvão mineral (1,6%). Com a geração de eletricidade da biomassa, pode-se considerar que ocorre uma compensação do carbono liberado na queima do material vegetal pela absorção desse elemento no crescimento das plantas. Entretanto, estudos indicam que as emissões de metano ($CH_4$) das hidrelétricas podem ser comparáveis às emissões de $CO_2$ das termelétricas.

MORET, A. S.; FERREIRA, I. A. As hidrelétricas do Rio Madeira e os impactos socioambientais da eletrificação no Brasil. **Revista Ciência Hoje**. V. 45, n° 265, 2009 (adaptado).

No Brasil, em termos do impacto das fontes de energia no crescimento do efeito estufa, quanto à emissão de gases, as hidrelétricas seriam consideradas como uma fonte

Ⓐ limpa de energia, contribuindo para minimizar os efeitos deste fenômeno.
Ⓑ eficaz de energia, tomando-se o percentual de oferta e os benefícios verificados.
Ⓒ limpa de energia, não afetando ou alterando os níveis dos gases do efeito estufa.
Ⓓ poluidora, colaborando com níveis altos de gases de efeito estufa em função de seu potencial de oferta.
Ⓔ alternativa, tomando-se por referência a grande emissão de gases de efeito estufa das demais fontes geradoras.

**Figura 23.** Questão 80 da prova azul de Ciências da Natureza do Enem 2011.

---

[6] Disponível em: <http://www.apoena.org.br/artigos-detalhe.php?cod=207>. Acesso em 20/08/2014.

**Tabela 15.** Percentual de escolha de cada uma das alternativas da questão (o gabarito é a letra D).

| 2011_q80 | Percentual de acerto (%) |
|---|---|
| A | 17,2 |
| B | 15,8 |
| C | 23,4 |
| **D** | **21,4** |
| E | 21,6 |

As curvas características da questão são mostradas na Figura 27a. O ajuste do modelo indica que a questão, que aparentemente tem sua solução no próprio texto, revelou-se muito difícil, com parâmetro de dificuldade 720, com uma discriminação adequada e parâmetro de acerto casual de 20%. Na Figura 27b, as curvas de percentual de escolha por distrator em função da nota revelam que até 500 pontos as escolhas são distribuídas uniformemente entre as alternativas. O distrator (C) é ligeiramente dominante até 550 pontos, e a partir deste valor a alternativa mais marcada é a resposta correta, (D). No entanto, pode-se observar um pequeno aumento nas escolhas da alternativa (B), "eficaz de energia", na nota 800.

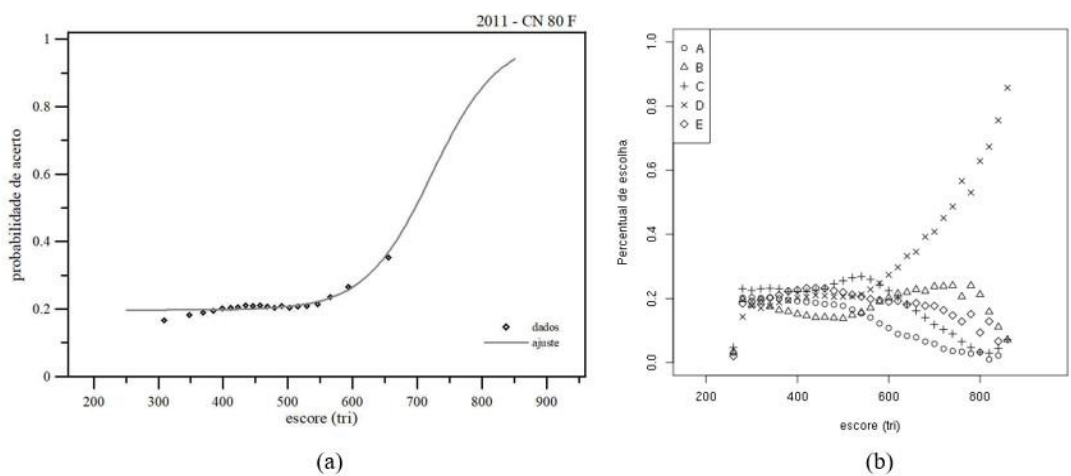

**Figura 24**. Questão 80 de 2011 (hidrelétrica). (a) Curvas características do item. (b) Percentuais de escolha de cada uma das alternativas em função da nota do aluno.

Esta questão é interessante, é conceitual e a resposta está contida no texto. Entretanto, sendo a matriz energética brasileira preponderantemente hidroelétrica, há a visão equivocada que esta não causa danos relevantes ao meio ambiente.

**As concepções sobre ótica geométrica e a questão 2012_Q64**

A questão 2012_Q64, apresentada na Figura 28, é uma questão conceitual que aborda o fenômeno da refração da luz quando esta passa da água para o ar.

Na Tabela 16, estão indicados os percentuais de escolha de cada uma das alternativas. Observa-se que cerca de 30% dos estudantes escolheram os distratores (B) e (D), alternativas que revelam que os concluintes do ensino médio possuem a noção equivocada de que, para conseguir enxergar, os olhos devem emitir raios luminosos [44;45]. A alternativa com maior percentual de escolha, de 28%, é o gabarito.

**QUESTÃO 64**

Alguns povos indígenas ainda preservam suas tradições realizando a pesca com lanças, demonstrando uma notável habilidade. Para fisgar um peixe em um lago com águas tranquilas o índio deve mirar abaixo da posição em que enxerga o peixe.

Ele deve proceder dessa forma porque os raios de luz

- **A** refletidos pelo peixe não descrevem uma trajetória retilínea no interior da água.
- **B** emitidos pelos olhos do índio desviam sua trajetória quando passam do ar para a água.
- **C** espalhados pelo peixe são refletidos pela superfície da água.
- **D** emitidos pelos olhos do índio são espalhados pela superfície da água.
- **E** refletidos pelo peixe desviam sua trajetória quando passam da água para o ar.

**Figura 28.** Questão 64 da prova azul de Ciências da Natureza do Enem 2012.

**Tabela 16.** Percentual de escolha de cada uma das alternativas da questão (o gabarito é a letra E).

| 2012_q64 | Percentual de acerto (%) |
|---|---|
| A | 23,2 |
| B | 19,4 |
| C | 18,8 |
| D | 10,6 |
| **E** | **27,6** |

Na Figura 29a, estão a curva característica do modelo e os dados empíricos. Esta curva indica que a questão é moderadamente difícil, com parâmetro de dificuldade 575 pontos, com grande discriminação e acerto casual de cerca de 15%. Na Figura 29b, estão os percentuais de escolha de cada uma das alternativas como função da faixa de nota. Observa-se que a resposta correta só é a mais escolhida para respondentes com mais de 570 pontos, e que a alternativa (A), que indica que os raios não se propagam em linha reta dentro da água, é a mais importante até ser ultrapassada pela resposta correta.

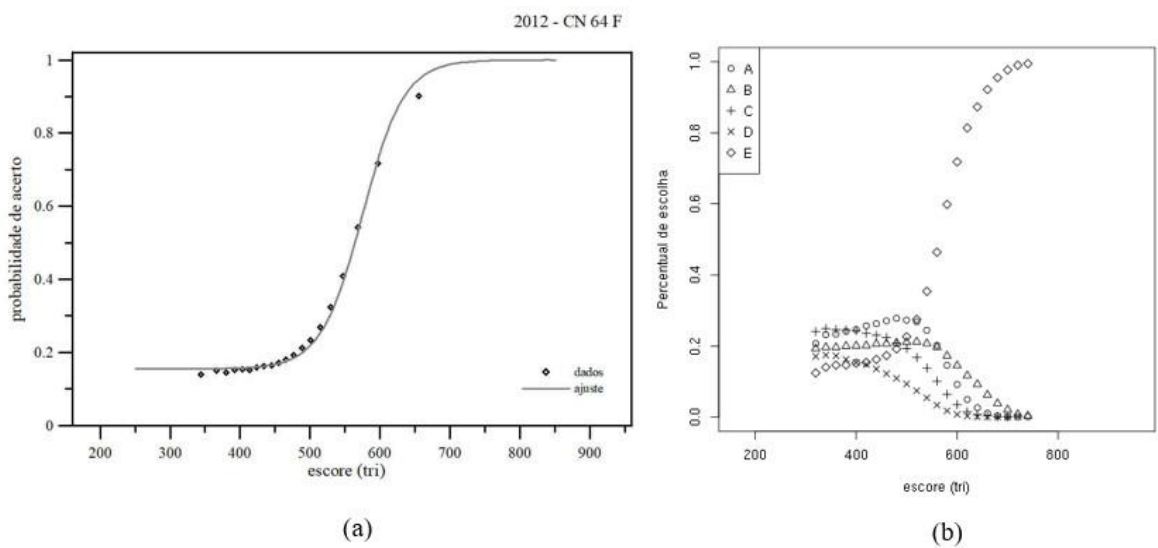

**Figura 29**. Questão 64 de 2012 (pesca). (a) Curvas características do item. (b) Percentuais de escolha de cada uma das alternativas em função da nota do aluno.

A questão proposta é bem apresentada, e o modelo geométrico para a luz é um tema muito presente na educação básica. O distrator mais escolhido revela uma incompreensão do mecanismo de propagação da luz que envolve a visão, o desconhecimento da refração da luz e indica a persistência de concepções não científicas também neste tema.

**Considerações Finais**

O Exame Nacional do Ensino Médio (Enem) vem se revelando como uma fonte relevante de dados para a avaliação do que é aprendido pelos estudantes concluintes do ensino médio [6;8;10;11;12].

Neste trabalho, apresentam-se os resultados de um conjunto de questões (num total de treze) das provas de Ciências da Natureza dos exames dos anos de 2009 a 2014. Essas questões foram classificadas como "questões de física", e apresentam concepções não científicas em seus distratores (as alternativas que não constituem o gabarito). Não foram consideradas aqui as questões relacionadas ao tema "fenômenos elétricos e magnéticos" dentre todos os objetos de conhecimento da Matriz de Referência.

Os dados analisados incluíram todas as respostas disponíveis na página do Inep para alunos que são concluintes do ensino médio no ano do exame; o número desses alunos varia entre 860 e 1370 mil. A numeração das questões refere-se à da prova azul. Os resultados apresentados correspondem à obtenção de dados percentuais e de parâmetros da TRI obtidos pelos autores.

A escolha pelos estudantes de alguns desses distratores revela que conceitos básicos de mecânica, fenômenos térmicos e ótica geométrica não são aprendidos pela maioria dos concluintes do ensino médio. Quase todas as questões escolhidas são conceituais, consideradas difíceis, normalmente discriminando estudantes com um desvio padrão ou mais acima da média. Há duas questões com discriminação negativa, o que não está de acordo com o modelo matemático utilizado para a TRI. Talvez o fato de existir um distrator vinculado a concepções não científicas muito fortemente presentes na estrutura cognitiva do estudante faça com que algumas das hipóteses da TRI não sejam inteiramente satisfeitas, resultando em desfavorecimento no desempenho de alunos com altas aptidões. Essa é uma hipótese que deveria ser avaliada por especialistas em psicometria. No entanto, a presença no exame de questões psicometricamente inadequadas possibilitou um estudo mais detalhado sobre o que os alunos aprenderam.

O resultado obtido revela que, apesar de todo o esforço desenvolvido na área de pesquisa em ensino de física desde os anos 1980, houve pouco impacto dos resultados no processo de aprendizagem. Mesmo com o conhecimento de que há dificuldades em modificar concepções presentes na estrutura cognitiva do aluno, os resultados são ainda muito impactantes e desanimadores.

Espera-se com este trabalho contribuir para uma reflexão sobre a importância da discussão desses dados nos cursos universitários, tanto de licenciatura (que forma professores para a educação básica) quanto de bacharelado (que forma professores para o ensino superior).

É sempre importante lembrar que (Ausubel apud Moreira [46])

> *"... o fator isolado mais importante que influencia a aprendizagem é o que o aluno já sabe; descubra isso e ensine-o de acordo ..."* (p.163)

Sem conhecimento do que o aluno já sabe, é difícil promover qualquer processo efetivo de ensino e aprendizagem.

**Referências**